\documentclass{pasa}%

\usepackage{framed}

\usepackage[authoryear]{natbib}
\bibpunct{(}{)}{;}{a}{}{,}
\usepackage{pdflscape}
\usepackage{array}

\usepackage{subcaption}
\usepackage{enumerate}

\let\chapter\section
\usepackage[lined,ruled,commentsnumbered]{algorithm2e}

\usepackage{tikz}
\usetikzlibrary{shapes.geometric, arrows, positioning, decorations.markings}

\title[PUMA]{PUMA: The Positional Update and Matching Algorithm}

\author[J.~L.~B.~Line]{J.~L.~B.~Line$^{1,2}$\thanks{{\href{malito:jline@student.unimelb.edu.au}{jline@student.unimelb.edu.au}}}
R.~L.~Webster$^{1,2}$,
B.~Pindor$^{1,2}$,
D.~A.~Mitchell$^{3,2}$ \and
C.~M.~Trott$^{4,2}$
\\
\\
\affil{$^1$The University of Melbourne, Melbourne, Australia}%
\affil{$^2$ARC Centre of Excellence for All-sky Astrophysics (CAASTRO)}
\affil{$^3$CSIRO Astronomy and Space Science (CASS), Epping, Australia}
\affil{$^4$International Centre for Radio Astronomy Research, Curtin University, Perth, Australia}
}

\jid{PASA}
\doi{10.1017/pas.\the\year.xxx}
\jyear{\the\year}

\usepackage{hyperref}
\hypersetup{
  colorlinks   = true, 
  urlcolor     = blue, 
  linkcolor    = blue, 
  citecolor    = black 
}

\begin{document}

\begin{abstract}
We present new software to cross-match low-frequency radio catalogues: the Positional Update and Matching Algorithm (PUMA). PUMA combines a positional Bayesian probabilistic approach with spectral matching criteria, allowing for confusing sources in the matching process. We go on to create a radio sky model using PUMA based on the Murchison Widefield Array Commissioning Survey, and are able to automatically cross-match~\textbf{$\sim98.5\%$} of sources. Using the characteristics of this sky model, we create simple simulated mock catalogues on which to test PUMA, and find that PUMA can reliably find the correct spectral indices of sources, along with being able to recover ionospheric offsets. Finally, we use this sky model to calibrate and remove foreground sources from simulated interferometric data, generated using OSKAR (the Oxford University visibility generator). We demonstrate that there is a substantial improvement in foreground source removal when using higher frequency  and higher resolution source positions, even when correcting positions by an average of $0.3'$ given a synthesized beam-width of~$\sim2.3'$.

\end{abstract}

\begin{keywords}
catalogue -- methods: statistical -- galaxies: statistics -- reionisation 
\end{keywords}

\maketitle

\section{Introduction}
\label{sec:intro}

Over the past decade, a new generation of low wide-field radio-frequency ($\leq\sim1\,$GHz) radio telescopes (e.g. \citealp[LOFAR,][]{VanHaarlem2013}; \citealp[MWA,][]{Tingay2012}; \citealp[PAPER,][]{Parsons2010}) has emerged that require fundamentally different calibration and imaging techniques to traditional radio astronomy. Gone are the days of a simple calibration pointing at a single point source~\citep[see][]{Smirnov2015}; with fields of 10s of degrees there are no isolated point sources, and instruments can become confusion limited in 10s of seconds~\citep{Bowman2012}. Indeed, new algorithms to include wide-field effects have been developed~\citep[e.g.][]{Rau2009} and calibration techniques utilising multiple calibrators from across the sky have been employed~\citep{Kazemi2013}.

Creating an all-sky and reliable catalogue with which to calibrate low radio-frequency astronomical data is then a necessary task. The ideal calibration catalogue would span multiple frequencies, providing a reliable spectral shape for each source. It would also be free from any ionospheric positional offsets; it could then be used to correct for ionospheric refraction in future observations~\citep[e.g.][]{Mitchell2008}. To improve our understanding of the radio sky, efforts are currently under way to create ever deeper surveys below $250\,$MHz in both the northern (\citealp[MSSS,][]{Heald2015}; \citealp[TGSS ADR1\footnote{the first alternative data release of the TIFR GMRT sky survey - see \url{http://tgssadr.strw.leidenuniv.nl/doku.php}},][in press]{Intema2016a}) and southern~\citep[GLEAM,][]{Wayth2015} hemispheres.

Each surveying instrument is limited in the frequencies it can access however, so to gain more frequency coverage, multiple catalogues must be combined. There are many examples of cross-matching techniques for radio wavelength data in the literature~\citep[e.g.][]{Kimball2008,Naylor2013,Fan2015}. %
Each method seeks to overcome the difficulty in matching sources found from surveys observed with varying instruments and frequencies. Not only does each telescope have its own resolution and sensitivity, but the morphology of each source may change with frequency. Furthermore, each catalogue employs its own source finding algorithm, which inherently has its own strengths and weaknesses. Sophisticated cross-matches are also prevalent in the optical literature~\citep[e.g.][]{Haakonsen2009,Pineau2011,Bilicki2016}, which use their own probabilistic positional matching. These cross-matches often focus on finding one single true cross-match between two particular catalogues however, and as such are often necessarily bespoke as to fold in known catalogue selection effects and to achieve the desired science. There are also generic cross-matching tools (such as provided by \texttt{SExtractor}~\citep{Bertin1996}), but of course these are designed to work with optical magnitudes, and so take work to feed radio wavelength catalogues into.

There is a further need for highly accurate radio sky models, for the current generation of low radio-frequency arrays attempting to measure the 21cm Hydrogen emission line during the Epoch of Reionisation (EoR) (e.g. MWA, LOFAR, PAPER). For these experiments, local galactic and extra-galactic radio sources act as foreground objects, masking the desired signal, and must be removed with exquisite precision~\citep[see][for reviews]{Furlanetto2006,Morales2009,Pritchard2011}. 
As ionospheric offsets are expected to scale with $\sim \lambda^2$~\citep[e.g.][]{Intema2009}, and resolution scales with $1/\lambda$, higher frequency observations should have higher positional accuracies. If high frequency instruments can be used to gain precise positional information, allowing accurate removal of these foregrounds, this has a direct bearing on the design and implementation of new instruments for studying the EoR, such as the upcoming SKA\_{}LOW telescope~\citep{Dewdney2013}. Using the red-shifted 21cm line over a range of frequencies allows a probe of spatial scales parallel to the line of sight, as well as over cosmic evolution. It is essential then to also accurately capture the spectral behaviour of foreground sources, as an incorrect subtraction in the frequency domain can affect any derived EoR signal.

The Positional Update and Matching Algorithm (PUMA) was created to meet the needs outlined above. With this software, an approach is developed that utilises both source position and spectral information as matching criterion. Positions can be matched through probabilistic cross-identification, as described in~\citealp{Budavari2008}. The desirable quality of the approach outlined in~\citealp{Budavari2008} is that it can easily be scaled to any number of catalogues. Spectral information can be used as a second identification criteria, assuming a spectral model. By focussing purely on low radio frequencies, emission through synchrotron processes can be assumed, allowing the use of a simple power-law model. PUMA has also been created to be as generic as possible, to facilitate an all-sky cross-match that can then have further constraints applied to for any particular science goal.

The rest of the paper is organised as follows. In \S\ref{sec:Bayes}, we outline the theory of Bayesian probabilistic cross-identification. In \S\ref{sec:PUMA}, we detail the functionality of PUMA, and in \S\ref{sec:PUMAcat} we use PUMA to create a cross-matched catalogue using real data. Using the outcomes of this cross-match, we create mock catalogues to test the accuracy of PUMA is \S\ref{sec:testPUMA}. In \S\ref{sec:PSanaly} we introduce the 2D Power Spectrum and test the effects of inaccurate catalogue positions in foreground subtraction when measuring the EoR signal.  We discuss our results in \S\ref{sec:discuss}.

\tikzstyle{initial1} = [rectangle, rounded corners, minimum height=1cm, minimum width=3cm, text centered, text width=8cm, draw=black, fill=yellow!30]
\tikzstyle{initial2} = [rectangle, rounded corners, minimum height=1cm, minimum width=3cm, text centered, text width=9.5cm, draw=black, fill=yellow!30]
\tikzstyle{algo1} = [rectangle, rounded corners,  minimum width=3cm, minimum height=1cm, text centered, draw=black, fill=cyan!25, text width=4.5cm]
\tikzstyle{algo2} = [rectangle, rounded corners,  minimum width=3cm, minimum height=1cm, text centered, draw=black, fill=cyan!25, text width=3.2cm]
\tikzstyle{algo3} = [rectangle, rounded corners,  minimum width=3cm, minimum height=1cm, text centered, draw=black, fill=cyan!25, text width=3.0cm]
\tikzstyle{algo4} = [rectangle, rounded corners,  minimum width=3cm, minimum height=1cm, text centered, draw=black, fill=cyan!10, text width=4.1cm,dashed]
\tikzstyle{eyeball} = [rectangle, rounded corners,  minimum width=3cm, minimum height=1cm, text centered, draw=black, fill=blue!30, text width=3.0cm]
\tikzstyle{eyeball} = [rectangle, rounded corners,  minimum width=3cm, minimum height=1cm, text centered, draw=black, fill=blue!30, text width=3.0cm]
\tikzstyle{accept} = [rectangle, rounded corners,  minimum width=3cm, minimum height=1cm, text centered, draw=black, fill=olive!15!green!45, text width=4.5cm]
\tikzstyle{reject} = [rectangle, rounded corners,  minimum width=3cm, minimum height=1cm, text centered, draw=black, fill=red!50, text width=4.0cm]
\tikzstyle{label} = [rectangle, rounded corners,  minimum width=1cm, minimum height=1cm, text centered, draw=white, fill=white, text width=1.0cm]

\tikzstyle{arrow} = [thick,->,>=stealth]
\tikzstyle{arrow_d} = [thick,->,>=stealth,dashed]

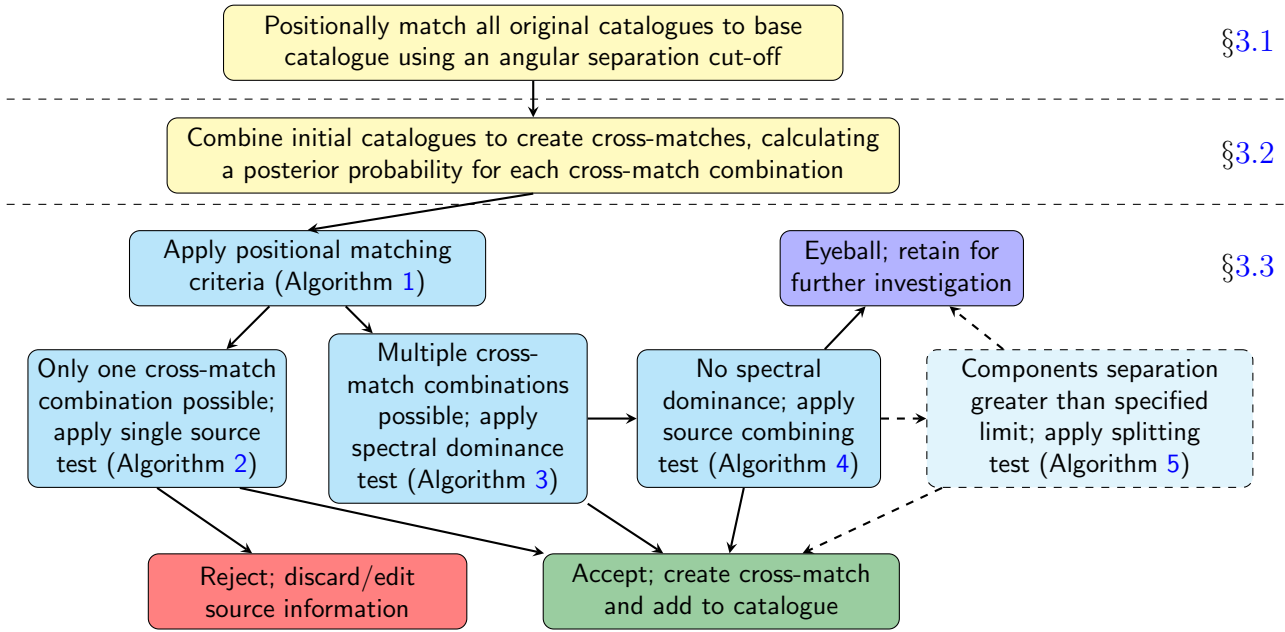
\begin{figure*}[t!]
\begin{tikzpicture}[node distance=1.5cm]
\node (stilts) [initial1, xshift=2cm] {\sffamily{Positionally match all original catalogues to base catalogue using an angular separation cut-off}};
\node (bayes) [initial2, below of=stilts] {\sffamily{Combine initial catalogues to create cross-matches, calculating a posterior probability for each cross-match combination}};

\node (algo1) [algo1, below of=bayes,xshift=-3cm] {\sffamily{Apply positional matching criteria (Algorithm~\ref{pos_crit_multi})}};
\node (eyes) [eyeball, below of=bayes, xshift=4.9cm] {\sffamily{Eyeball; retain for further investigation}};
\node (algo2) [algo2, below of=algo1, xshift=-2cm, yshift=-0.5cm] {\sffamily{Only one cross-match combination possible; apply single source test (Algorithm~\ref{pos_crit_single})}};
\node (algo3) [algo2, below of=algo1, xshift=2cm, yshift=-0.5cm] {\sffamily{Multiple cross-match combinations possible; apply spectral dominance test (Algorithm~\ref{spec_dom_crit})}};
\node (algo4) [algo3, below of=algo1, xshift=6cm, yshift=-0.5cm] {\sffamily{No spectral dominance; apply source combining test (Algorithm~\ref{comb_crit_test})}};
\node (algo5) [algo4, below of=algo1, xshift=10.4cm, yshift=-0.5cm] {\sffamily{Components separation greater than specified limit; apply splitting test (Algorithm~\ref{split_crit_test})}};
\node (accept) [accept, below of=algo4, xshift=-0.5cm, yshift=-0.8cm] {\sffamily{Accept; create cross-match and add to catalogue}};
\node (reject) [reject, below of=algo2, xshift=2cm, yshift=-0.8cm] {\sffamily{Reject; discard/edit source information}};

\node (s1) [label, xshift=11.5cm] { \large{\S\ref{int_match}} };
\node (s2) [label, yshift=-1.5cm, xshift=11.5cm] { \large{\S\ref{subsec_BayMatch}} };
\node (s3) [label, yshift=-3.0cm, xshift=11.5cm] { \large{\S\ref{match_crit}} };

\draw[dashed] (-5.0,-0.75) -- (12.0,-0.75);
\draw[dashed] (-5.0,-2.15) -- (12.0,-2.15);

\draw [arrow] (stilts) -- (bayes);
\draw [arrow] (bayes.south) -- (algo1.north);
\draw [arrow] (algo1) -- (algo2);
\draw [arrow] (algo1) -- (algo3);
\draw [arrow] (algo3) -- (algo4);
\draw [arrow] (algo4) -- (eyes);
\draw [arrow] (algo2.south) -- (reject);
\draw [arrow] (algo2.320) -- (accept);
\draw [arrow] (algo3) -- (accept);
\draw [arrow] (algo4) -- (accept);

\draw [arrow_d] (algo4) -- (algo5);
\draw [arrow_d] (algo5) -- (eyes);
\draw [arrow_d] (algo5) -- (accept);

\end{tikzpicture}
\caption{\sffamily{The steps and outcomes of the matching process are shown. Yellow boxes represent steps with no criteria applied, cyan represent criteria being applied, and all other colours represent final points. Each cyan box refers to a specific Algorithm, as detailed in Algorithms~\ref{pos_crit_multi}-\ref{split_crit_test}. The section labels on the right refer to \S\ref{int_match},~\ref{subsec_BayMatch}, and~\ref{match_crit}, which detail each step. Each section is performed by a separate script.}}
\label{flow}
\end{figure*}
\section{Bayesian positional cross-matching}
\label{sec:Bayes}
In a Bayesian analysis, an hypothesis $H$ can be related to its complementary hypothesis $K$ through the Bayes Factor
\begin{equation}
B(H,K|D) = \frac{P(H|D)/P(H)}{P(K|D)/P(K)} ,
\end{equation}
where $D$ is some measurement set. For this application, $D$ is a set of source positions from multiple catalogues, $H$ is the situation where each catalogue is reporting the same astrophysical source, and $K$ is where they are not. The larger the value of $B(H,K|D)$ then, the stronger the support for the hypothesis $H$, which in this case indicates a good cross-match. When matching $n$ catalogues, it can be shown \citep[see][for further details]{Budavari2008} that the Bayes Factor is given by:
\begin{equation}
\label{bayes_fact}
B = 2^{n-1} \frac{\prod w_i}{\sum w_i} \exp \left( -\frac{\sum_{i<j} w_iw_j\psi^2_{ij}}{2\sum w_i} \right) ,
\end{equation}
where $\psi^2_{ij}$ is the angular separation between sources in the $i^{th}$ and $j^{th}$ catalogues, and $w_i$ is the weighting for the $i^{th}$ catalogue. This is given by $w = 1/\sigma^2$, where $\sigma^2$ is the astrometric precision. This is calculated as
\begin{equation}
\label{errors}
\sigma^2 = \sigma_{\textrm{RA}}^2 + \sigma_{\textrm{Dec}}^2,
\end{equation}
where $\sigma_{\textrm{RA}}$, $\sigma_{\textrm{Dec}}$ are the errors on right ascension and declination, respectively. These errors are usually quoted directly in each source catalogue. 
The Bayes Factor can be related to the posterior probability through
\begin{equation}
P(H|D) = \left[ 1 + \frac{1 - P(H)}{BP(H)} \right]^{-1} .
\label{posterior}
\end{equation}

For multiple catalogues, the prior may be calculated through
\begin{equation}
P(H) = \frac{\nu_\star}{\prod \nu_i} \left(\frac{\Omega_{overlap}}{4\pi} \right)^{(1-n)} ,
\label{prior}
\end{equation}
where the scaled full sky number of sources in each catalogue $\nu$ is give by $\nu_i = 4\pi N_i/ \Omega_i$, with $N_i$ the number of sources in the sky area $\Omega_{i}$. $\nu_\star$ is the scaled full sky number of sources in the final matched catalogue, with 
$\Omega_{overlap}$ the region of sky where all matched catalogues overlap. This calculation simply accounts for the source density of each catalogue and how much of the sky all catalogues cover as a way of estimating the prior. The true $P(H)$ depends also on the selection effects of each catalogue; for example if one catalogue is more sensitive to diffuse emission, the final catalogue may be biased toward brighter flux densities as it is able to detect more emission. There is no simple way for the user to enter these subtle selection effects however, and in the low radio-frequency regime each catalogue should see similar astrophysical skies, hence this simple prior is retained. Future releases could potentially include a way to incorporate custom selection effects.

\section{PUMA}
\label{sec:PUMA}
PUMA is an openly available code\footnote{PUMA is stored in a repository here: \url{https://github.com/JLBLine/PUMA}} which is free to use. The flow of the matching algorithm is shown in Figure~\ref{flow}. The following sections expand upon the methodology of each step.

As explained in \S~\ref{sec:Bayes}, $P(H|D)$ gives the posterior probability of a single match, base purely on the positions, positional errors and source densities of each respective catalogue. However, this does not take in to account the resolution of the differing instruments and surveys used to create each catalogue. It may be the case that a catalogue with lower resolution is indeed describing the same astrophysical source, but averaging over many components, thus measuring a combined flux density and position. Comparing these two catalogues may then yield a very low probability that they are exactly the same source. To account for this, PUMA uses both positional and spectral data to evaluate matches, and allows multiple sources from the same catalogue to be considered in a single cross-match. In this way any number of catalogues can be matched, but at the cost of using a designated catalogue as a base to cross-match to. In the following section, the following terminology is used:

\begin{list}{}{}
\item[\textbf{\textit{source}}] - a single entry in a catalogue
\item[\textbf{\textit{base catalogue}}] - the catalogue upon which all positional cross-matching is performed
\item[\textbf{\textit{cross-match combination}}] - defined as a cross-match of sources including only one source from each catalogue
\item[\textbf{\textit{repeated catalogue}}] - if more than one source from a single catalogue is matched to a base source, that catalogue is termed a repeated catalogue
\end{list}

The following steps broadly describe the operations carried out by PUMA; they are elaborated upon in \S\ref{int_match},~\ref{subsec_BayMatch}, and~\ref{match_crit} respectively:

\begin{enumerate}[1.]
  \item Positionally cross-match all catalogues individually to a base catalogue to some user defined cut-off separation. Retain all cross-matches for each base source.
  \item For each base source, use the cross-matched tables to create every possible cross-match combination including the base source. Calculate $P(H|D)$ for each cross-match combination.
  \item Apply positional and spectral criteria to each base source and set of matched combinations to identify the best cross-match combination.
\end{enumerate}

Each of these steps is carried out by the scripts \texttt{cross\_{}match.py}, \texttt{calculate\_{}bayes.py}, and \texttt{create\_{}table.py}, respectively. This allows the user greater flexibility in modifying parameters at any stage.

\subsection{Initial Positional Match}
\label{int_match}
\begin{figure}[t!]
\centering
\includegraphics[width=\columnwidth]{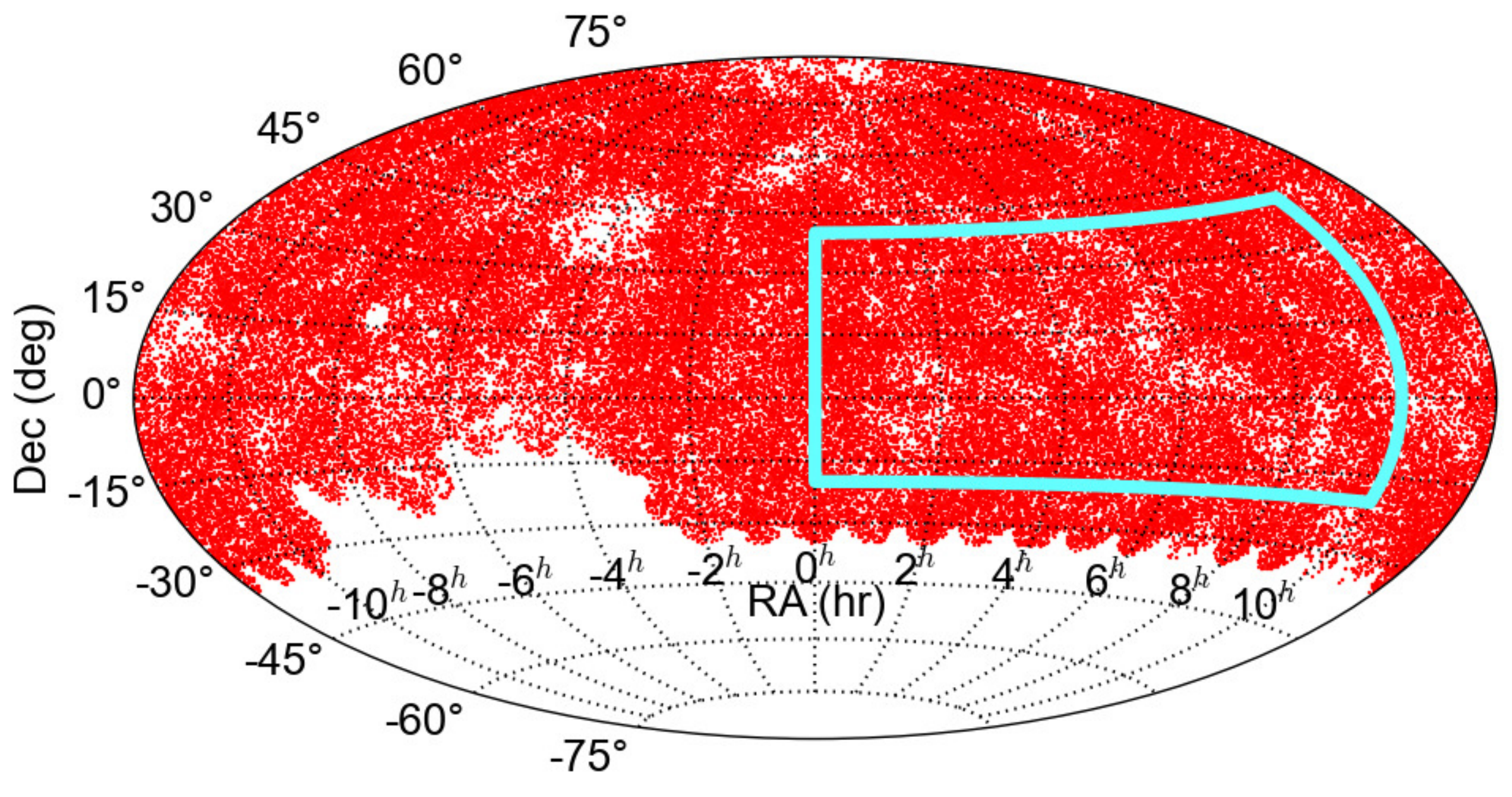}
\caption{\sffamily{
All sources in the VLSSr~\citep{Lane2014} catalogue are plotted. To calculate the source density of the catalogue, \texttt{cross\_{}match.py} takes given RA and Dec bounds, and counts the number of sources within that area. In this example, the limits are represented by the cyan lines. It is left to the user to pick an area that will give a representative source density of the entire catalogue. For example, if too small an area, or a particularly under-dense area such as that at $RA,\delta = 4^\textrm{h},40^\circ$, is selected, an unrepresentative source density will be calculated.
\label{source_dens}
}}
\end{figure}

The script \texttt{cross\_{}match.py} supports the standard FITS and VOTable formats. Each input catalogue is cross-matched with a designated base catalogue. The final matched catalogue will have a similar source density to the base catalogue, so the user should select the base catalogue that suits their needs. \texttt{cross\_{}match.py} performs two functions. Firstly, relevant information from each catalogue is formatted in a standard way. The user must specify the column names and units of the catalogue, which are then converted and saved in to a `simple' fits table. The required information taken from each catalogue is: Source Name; $RA\,$J2000$\,$(deg); Error on $RA\,$(deg); $\delta\,$J2000$\,$(deg); Error on $\delta\,$(deg); Flux density$\,$(Jy); Error on Flux density$\,$(Jy). Optionally, the user can supply: Major Axis$\,$(deg); Minor Axis$\,$(deg); Position Angle$\,$(deg); Flags; Field IDs. The two latter columns are included for the information of the user, but are currently unused by PUMA. These standard tables are then used to perform a positional cross-match within a given cut-off distance using the STILTS\footnote{STILTS documentation - \url{http://www.star.bris.ac.uk/~mbt/stilts/}} package~\citep{Taylor2006}. STILTS is the command line version of the cross-matcher used in TOPCAT. \texttt{cross\_{}match.py} uses the information given by the user to create a matched catalogue, where every possible match of the specified catalogue to a source in the base catalogue is saved. Each catalogue is matched to the base catalogue individually, so any number of combinations of catalogues can be used by \texttt{calculate\_{}bayes.py} (see \S\ref{subsec_BayMatch}). \texttt{calculate\_{}bayes.py} needs two further things; the source density of each catalogue, and the sky coverage of each catalogue, in order to calculate $P(H)$ (Eq.\ref{prior}). \texttt{cross\_{}match.py} internally calculates the sky coverage of each catalogue, fully taking into account the continuous nature of RA co-ordinates. To calculated the source density, the user specifies an area on the sky, bound by lines of RA and Dec for each catalogue. \texttt{cross\_{}match.py} then simply counts the number of sources within this lune. These data are stored in the meta-data of both the individual `simple' tables and the final matched table in a standard way, allowing \texttt{calculate\_{}bayes.py} to automatically read the data. The sky lune to measure the source density within is left to the user's discretion; an example is shown in Figure~\ref{source_dens}.

\subsection{Bayesian Match Calculation}
\label{subsec_BayMatch}
The theory outlined in \S\ref{sec:Bayes} is implemented in \texttt{calculate\_{}bayes.py}. This script uses any number of the matched tables created by \texttt{cross\_{}match.py}, combines them, and then calculates a posterior probability for every possible cross-match combination involving each base source.

To calculate the posterior probability (Eq.\ref{posterior}) for each combination, $B$ (Eq.\ref{bayes_fact}) and $P(H)$ (Eq.\ref{prior}) must be calculated. For the weights in $B$, the quoted positional errors from each catalogue are used as shown in Eq.\ref{errors}.The rest of the calculation of $B$ is straight forward. To calculate $P(H)$, the number of sources scaled to a full sky coverage for each catalogue $\nu_i$, as well as that of the final matched catalogue $\nu_\star$, must be known. These values have already been calculated by \texttt{cross\_{}match.py}, and are read in automatically. Once a combination of potential matches has been formed, \texttt{calculate\_{}bayes.py} identifies the present catalogues, uses the sky coverages measured by \texttt{cross\_{}match.py} to calculate $\Omega_{overlap}$ (which may different depending on which catalogues are involved in the match), and uses the applicable $\nu_i$ to calculate $P(H)$. $\nu_\star$ is assumed to be the source density of the base catalogue.

\subsection{Matching Criteria}
\label{match_crit}
The information generated by \texttt{calculate\_{}bayes.py} is used by \texttt{make\_{}table.py} to create a final matched catalogue, by applying the steps shown in the lower section of Figure~\ref{flow}. 

The first criteria applied determines which cross-match combinations are deemed as positionally possible. When running \texttt{make\_{}table.py}, the user supplies two variables that dictate what PUMA defines as a possible cross-match combination: $P_u$ and $\theta_r$. $P_u$ is a positional probability threshold; if $P(H|D) > P_u$, the cross-match combination is deemed possible regardless of the separation between the individual matched sources. $\theta_r$ is the resolution of the base catalogue (which is usually set to the FWHM of the instrument response of the catalogue). As noted earlier, catalogues with higher resolution may resolve multiple components within the FWHM of a lower resolution catalogue. This may yield low posterior probabilities even though these matches are true; the original assumption in the bayesian treatment is that each catalogue has one true match, and as such doesn't take into account blending of sources at lower resolutions. PUMA therefore considers any cross-match combination where all sources lie within the resolution of the base catalogue plus error to be possible. Explicitly, any source lying within an error ellipse defined as
\begin{equation}
\left(\frac{\Delta\theta_{\textrm{RA}}}{ (\theta_r / 2) + \sigma_{\textrm{RA}}}\right)^2 + \left(\frac{\Delta\theta_{\textrm{Dec}}}{ (\theta_r / 2) + \sigma_{\textrm{Dec}}}\right)^2 \leq 1,
\label{eq_errEllip}
\end{equation}
where $\Delta\theta_{\textrm{RA}}$, $\Delta\theta_{\textrm{Dec}}$ are the angular offsets of the source from the base source in Right Ascension and Declination respectively, is considered possible, regardless of the positional probability derived by the bayesian treatment. These positional criteria are initially applied by \texttt{calculate\_{}bayes.py} (\S\ref{subsec_BayMatch}). Any matched source from a non-repeated catalogue will be present in all cross-match combinations; as such, these sources are subjected to the above test during the initial calculation of $P(H|D_i)$ for all cross-match combinations associated with a single base source. If all $P(H|D_i) < 0.5$, matched sources from a non-repeated catalogue are tested using Eq.~\ref{eq_errEllip}. If they fail, the matched sources are discarded, and $P(H|D_i)$ is recalculated.

\texttt{make\_{}table.py} then uses Algorithm~\ref{pos_crit_multi} to apply these positional tests to matched sources from repeated catalogues, to discard any unlikely cross-match combinations. (All algorithms used in \texttt{make\_{}table.py} can be found in Appendix~\ref{app_Alg}). 

The successful candidate cross-match combinations are then passed through the following criterion. PUMA splits the matching in to three main regimes; these are explained in more detail in \S\ref{subsec_IsoMatch}, \ref{subsec_DomMatch}, and \ref{subsec_MulMatch}:
\begin{enumerate}[1.]
\item \texttt{isolated}: These are matches where only one cross-match combination is present. These either occur naturally, or when all other cross-match combinations fail the positional matching criteria. PUMA will accept an isolated match if $P(H|D)$ is above some probability threshold, or if the spectral data of the matched sources fit a power law spectral model. If neither of these criteria are met, the match is flagged as rejected.
\item \texttt{dominant}: If there are multiple sources from a repeated catalogue that are deemed as possible, each cross-match combination is tested for dominance. This is defined as when the residuals of a fit to a power law spectral model of one combination of sources are at least three times smaller than all other combinations, and the probability of that one combination is larger than all other combinations.
\item \texttt{multiple}: If no dominant source is found, the multiple sources from a repeated catalogue are combined in to a single source. These new combined flux densities are then fit with a power law. If the fit is good, the match is accepted with the new combined flux density and positions, generated by combining the sources in the same catalogue. If the combined source fails the spectral test, it is flagged for visual inspection.
\end{enumerate}

\subsubsection{Isolated Matches}
\label{subsec_IsoMatch}
If a base source is labelled as \texttt{isolated} by Algorithm~\ref{pos_crit_multi}, it is passed to Algorithm~\ref{pos_crit_single}. If $P(H|D) \geq P_u$, the cross-match combination is accepted without further investigation. If $P(H|D) < P_u$, all sources in the cross-match combination are tested using Eq.~\ref{eq_errEllip}. If all sources pass, the spectral information is interrogated. The flux densities $\boldsymbol{f}$ at frequencies $\boldsymbol{\nu}$ are tested by fitting a linear model 
\begin{equation}\label{eq_specMod}
	\ln(\boldsymbol{f}) = \alpha \ln(\boldsymbol{\nu}) + \beta
\end{equation}
using weighted least squares. This is done using the python package \texttt{statsmodels}
\footnote{statmodels documentation - \url{http://statsmodels.sourceforge.net/devel/index.html}}.
The residuals of these fits are then investigated to ascertain the deviation of the data from a linear fit. The goal of this spectral fit is to identify large deviations from linearity in $\log$ space; it is designed to allow for some curvature of the data. Over small enough frequency ranges, most spectra are expected to follow the linear model in Eq.~\ref{eq_specMod}, but varying degrees of curvature may be inherent~\citep[see][and references therein]{Torniainen2008}. Given that there are a limited number of low radio-frequency catalogues to match, the number of data points are often low. This rules out easily using models that account for curvature, as the number of parameters to fit quickly outnumbers the number of data points. As such, the spectral test detailed here has been designed to be simple and robust, and tunable by the user to meet any desired criteria as much as possible.

To test the `goodness' of the fit, the reduced chi-squared value $\chi^2_{red}$ is typically inspected e.g.~\citealp{Hogg2010}. This value sums the residuals of the fit in units of the variance of the data, and scales for the complexity of the fitted model as
\begin{equation}
\label{chi_resids}
\chi_{red}^2 = \frac{1}{K} \sum^n_{i=1} \left( \frac{ \ln(f_i) - F_i}{\sigma_i} \right)^2 ,
\end{equation}
where $n$ is the number of matched catalogues, $\sigma_i$ is quoted error on flux density, $F_i$ is the modelled fit of $\ln(f_i)$, and $K = (n - p)$, with $p$ being the number of parameters set ($p$=2 for a linear fit). As a general rule, a result of $\chi^2_{red}<=2$ is considered a good fit, as the model lies within twice the observed variance of the data. However, as explored in~\citealp{Andrae2010}, $\chi^2_{red}$ comes with its own uncertainty which grows as the number of data points reduces. Further to this, $\chi^2_{red}$ is reliant on the errors on each observations being drawn from a Gaussian distribution. Given that each data set contains data points subject to differing error analyses, the extent to which this is true is difficult to ascertain. Practically, it has been found that $\chi^2_{red}$ works well at classifying the fit at low flux densities, where the quoted errors are comparable to the magnitude of the residuals. Conversely, a data set that displays a similar curvature in log space at a much higher flux density yields a far larger $\chi^2_{red}$ value, as the errors are small compared to the residuals of the linear fit. Using $\chi^2_{red}$ with some cut-off threshold then tends to reject the brightest sources, which are the best sources to calibrate on~\citep[e.g.][]{Mitchell2008}. To include the brightest sources, a second residual metric, $\epsilon$, is defined as
\begin{equation}
\epsilon = \frac{1}{n} \sum^n_{i=1}\left( \frac{|f_i - \exp(F_i)|}{f_i} \right).
\end{equation} 

This residual is designed to test the deviation of a fit from the data in units of the observed value\footnote{Note that this residual is performed in the natural units of the flux density, to ensure all flux density values are positive and a correct scaling is found.}. By scaling $\epsilon$ by the number of data points, $\epsilon$ describes the total deviation of the data points from a linear fit as a fraction of the magnitude of the data points. This metric performs well at high flux densities, where the residuals of the fit are small compared to the magnitude of the flux density, but poorly at low flux densities where the residuals are comparable to or larger than the flux densities. $\chi_{red}^2$ and $\epsilon$ are then complimentary metrics. 

Both are used with cut-off thresholds $\epsilon_u$, $\chi_{red,u}^2$ in all spectral tests to accept a cross-match combination. These thresholds can be adjusted by the user to suit their needs. For an \texttt{isolated} match that has $P(H|D) < P_u$ but satisfies Eq.~\ref{eq_errEllip}, if either  $\epsilon^2 \leq \epsilon^2_u$ or $\chi_{red}^2 \leq \chi_{red,u}^2$, the cross-match combination is accepted. Otherwise, the information is stored and labelled as a \texttt{reject}.

\subsubsection{Dominant Matches}
\label{subsec_DomMatch}
If multiple cross-match combinations are deemed possible by Algorithm~\ref{pos_crit_multi}, they are subjected to a `dominance' test by Algorithm~\ref{spec_dom_crit}. Some sources from a repeated catalogue may be confusing sources that are not associated with the base catalogue source, even though they lie within the resolution of the base source. To test if one single cross-match combination is the correct match, the spectrum of each cross-match combination is fit using the model described by Eq.~\ref{eq_specMod}. If the residuals of the fit to one combination is 3 times less than all other fits, it is flagged as possibly dominant. Another criterion is demanded, being
$P(H|D_i) \geq P_u$ and $P(H|D_{i \neq j} < P_l$), where $D_i$ is the positional data of the possibly dominant cross-match combination, and $D_{i \neq j}$ the data of all other combinations. $P_l$ is again a threshold set by the user. These criteria guarantee that the chosen combination has a best fitting spectrum, as well as the most likely positional information. If a cross-match combination passes this test it is accepted and labelled as \texttt{dominant}.

\subsubsection{Multiple Matches}
\label{subsec_MulMatch}
A group of cross-match combinations that have no clear \texttt{dominant} combination are passed on to Algorithm~\ref{comb_crit_test} for a `combination' test. It is possible that sources from a repeated catalogue describe components that are unresolved by the base catalogue. Algorithm~\ref{comb_crit_test} tests this by combining the flux density of the sources from the repeated catalogues. It then tests the new spectral data by calculating both $\epsilon$ and $\chi^2_{red}$ of the new data. If the new combination has residuals under the residual thresholds, the new cross-match combination is accepted and labelled as a \texttt{multiple} match\footnote{There is a point to note with this test. If only one catalogue is matched to the base catalogue, there are only two data points to consider in a fit; a fit to two data points cannot have any residuals. Currently, for two data points, both $\chi^2_{red}$ and $\epsilon$ default to zero. This means for a repeated catalogue match, a spectral dominance test (Algorithm~\ref{spec_dom_crit}) will never find a difference between the residuals, and will always pass on to the combined source test (Algorithm~\ref{comb_crit_test}). Combining the sources will still give residuals of zero, so this case will always pass the combination test, and be given an updated weighted position.}. For each repeated catalogue, a new combined position is found through a weighting scheme described by
\begin{equation}
\label{simp_weight}
w_i = \frac{f_i}{\sum^n_i f_i} , 
\end{equation}
where $n$ is the number of repeated catalogue sources, and $f$ the flux density of each source. These weights are applied to create the new $RA$ position and error as
\begin{equation}
\label{new_RA}
RA_{\textrm{new}} = \sum^n_i w_iRA_i \quad , \quad  \sigma_{\textrm{new}}^2 = \sum^n_i (w_i\sigma_i)^2 ;
\end{equation}
this is similarly applied for $\delta$. If the combining test fails, the group source information is labelled as an \texttt{eyeball}. 

\textit{Splitting test} - combining sources as described above confines the output catalogue to the resolution of the base catalogue. \texttt{make\_table.py} also comes with an option to `split' the combined sources, which is handled by Algorithm~\ref{split_crit_test}. If the sources from each repeated catalogue are separated by a distance greater than some user specified cut-off, $d_{split}$, PUMA will attempt to split the combined source in to components. If there is more than one catalogue with repeated sources, PUMA currently requires that each repeated catalogue have the same number of sources matched, simplifying the matching of these repeated sources. The repeated sources from each repeated catalogue are matched by distance, to create new cross-match combination components. The flux density from each catalogue that had only one source matched is then split in to components to match these new cross-match combination components, based on the following weighting scheme. For each catalogue $m$ of $n$ repeated sources, weights are created and averaged as
\begin{equation}
w_k = \frac{1}{m} \sum^m_j w_{j,i}\quad,
\label{avg_weight}
\end{equation}
where $w_{j,i}$ is the $w_i$ (Eq~\ref{simp_weight}) weight in the $j^{th}$ catalogue. The flux density of each single catalogue match source is then split as
\begin{equation}
\mathbf{f_k} = f_s \mathbf{w_k}
\end{equation}
where $f_s$ is the flux density of the source, and $\mathbf{ w_k }$ is a vector of length $n$ of weights $w_k$. Applying the weights in this manner uses all the information from the repeated catalogues to create an accurate spectral energy distribution (SED) for each component.

Once an SED has been created for each component, they are spectrally tested as in Algorithm~\ref{spec_dom_crit}. If all components pass the spectral test, the sources are accepted. Each component is given a name based on the position of the original base source, with a letter appended to distinguish between components. If one or more components fail the spectral test, the source is flagged to investigate by eye.

\subsection{Final Matched Table}
\label{output_table}
Once \texttt{make\_{}table.py} has applied the algorithms described in \S\ref{match_crit}, a list of accepted sources is left. These are output to a either a FITS or VOTable, the contents of which are described in Table~\ref{cats_output}. Two positions are reported; that of the original base source, and an `updated' position. The user supplies a ranking of all matched catalogues. The position of the highest ranked source in a cross-match combination is then reported as this updated position. This ranking can be set to the needs of the user, but the highest rank would usually be given to the catalogue with the highest angular resolution and  least affected by ionospheric effects. If this highest ranked matched source was created by combining sources as described in \S\ref{subsec_MulMatch}, the flux density weighted centroid given by Eq.~\ref{new_RA} is reported.

As described in Eq.~\ref{eq_specMod}, a model of $ \ln(\boldsymbol{f}) = \alpha \ln(\boldsymbol{\nu}) + \beta $ is fit to each combination. When this fit is performed, the frequencies are entered in MHz and the flux densities in Jy. When performing the linear fit, the package \texttt{statsmodels} also outputs the standard error of the fitted parameters $\sigma^2_\beta,\sigma^2_\alpha$. The spectral index and error, $\alpha$ and $\sigma_{\alpha}$ are both reported in the final matched catalogue. If desired by the user, the fit is used to extrapolate a flux density and error at the frequency of the base catalogue through the equations
\begin{equation}
f_{ext} = e^\beta\nu^\alpha \quad , \quad \sigma_f^2 = \frac{1}{f^2_{ext}} \left( \sigma^2_\beta + \sigma^2_\alpha[\ln(\nu)]^2 \right)
\label{new_flux}
\end{equation}
where again the frequency is in MHz. 

The final table only includes sources that were accepted. For all base sources classified as \texttt{reject} or \texttt{eyeball}, all possible cross-match combinations are printed out to a text file, which can be used for further investigation. If selected, PUMA will print out a summary of the matching statistics to a text file. The PUMA package also comes with a plotting script, \texttt{plot\_{}outcomes.py}, which has multiple search criteria in built (for examples see Figures~\ref{fig:goodisolated}-\ref{fig:eyeball}). For detailed information on the running of, the outputs, and plotting capabilities of PUMA, refer to the online documentation.

\begin{table*}[t]
\renewcommand{\arraystretch}{1.2}
\caption{Details of the content of the final matched catalogue output by \texttt{make\_{}table.py}}
\begin{center}
	\begin{tabular*}{\textwidth}{p{4cm} p{12.5cm}} 
	\hline
	\textbf{Column Name} & \textbf{Description} \\
	\hline \hline
	\texttt{Name} & A name of the source based on the updated J2000 coordinates and a prefix supplied by the user \\
	\hline
	\texttt{*base*\_{}name} & The name of the source from the original base catalogue. Replace *base* with whatever prefix was chosen during creation \\
	\hline
	\texttt{updated\_{}RA\_{}J2000, updated\_{}DEC\_{}J2000, e\_{}updated\_{}RAJ2000, e\_{}updated\_{}DECJ2000} & The new J2000 RA and Dec given to the source (deg), chosen from a preferred catalogue as selected by the user. In the case of a combination of sources, the position and errors are calculated through Eq~\ref{new_RA} \\
	\hline
	\texttt{original\_{}RA\_{}J2000, original\_{}DEC\_{}J2000, e\_{}original\_{}RAJ2000, e\_{}original\_{}DECJ2000} & The original base catalogue source position and error (deg) \\
	\hline
	\texttt{S\_{}*freq*, e\_{}S\_{}*freq*} & The flux density and error reported by each catalogue (Jy) for a match, where *freq* is the frequency of the observation in MHz. \\
	\hline
	\texttt{Match\_{}type} & The type of the match as defined in \S\ref{match_crit}; either \texttt{isolated}, \texttt{dominant} or \texttt{multiple}. \\
	\hline
	\texttt{SI, e\_{}SI} & The spectral index and associated error to a weighted linear fit to the data ($\alpha$ in Eq.~\ref{eq_specMod}) \\	\hline
	\texttt{Intercept, e\_{}Intercept} & \textit{Optional:} The intercept and associated error to a weighted linear fit to the data ($\beta$ in Eq.~\ref{eq_specMod}) \\
	\hline
	\texttt{S\_{}*freq*\_{}ext, e\_{}S\_{}*freq*\_{}ext} & \textit{Optional:} The flux density (Jy) at the frequency of the base catalogue, as extrapolated by the parameters of the linear fit. These are calculated through Eq.~\ref{new_flux}. \\
	\hline
	\texttt{Number\_{}cats} & The number of catalogues in the cross-match combination, and hence the number of data points available to fit models to. \\
	\hline
	\texttt{Number\_{}matches, Retained\_{}matches} & \textit{Optional:} The number of possible cross-match combinations found by the positional matching criteria performed by \texttt{cross\_{}match.py} and \texttt{calculated\_{}bayes.py}, and the number retained after applying Algorithm~\ref{pos_crit_multi}. \\
	\hline
	\texttt{Match\_{}stage} & \textit{Optional:} The final decision point or data with which PUMA chose the cross-match combination; either \texttt{position} if matched purely by position, \texttt{spectral} if the spectral data was used, \texttt{combine} if any sources were combined or \texttt{split} if any sources were split. As an \texttt{isolated} source can be accepted through either \texttt{position} or \texttt{spectral}, this helps distinguish the two cases for any statistical analysis. \\
	\hline
	\texttt{Chi\_{}sq\_{}red} & \textit{Optional:} Returns the $\chi^2_{red}$ value for the fit \\
	\hline
	\texttt{*matched-cat*\_{}names} & \textit{Optional:} Returns a string of the names of all the matched catalogue sources. Replace *matched-cat* with whatever catalogue name was supplied by the user \\
	\hline \hline
	\end{tabular*}
\end{center}
\label{cats_output}
\end{table*}

\newcolumntype{P}[1]{>{\centering\arraybackslash}p{#1}}
\newcolumntype{M}[1]{>{\centering\arraybackslash}m{#1}}
\begin{table*}
\renewcommand{\arraystretch}{1.2}
\centering
\caption{\textsf{General characteristics of the base and cross-matching catalogues. The quoted beam widths for MWACS and MRC and indicative only, as they vary across the sky and with frequency.}}
	\begin{tabular}{P{1.6cm} P{1.4cm} P{1.5cm} P{1.8cm} P{2cm}}
	\hline
	\textbf{Catalogue} & \textbf{Frequency (MHz)} & \textbf{Beam width} & \textbf{Number of sources}  & \textbf{Image $rms$ (mJy)} \\
	\hline \hline
	VLSSr & 74 & 75$''$ & 92,696 & 100  \\
	MWACS & 180 & $\sim$3$'$ & 14,110 & 30 \\
	MRC & 408 & $\sim$2.7$'$ & 12,141 & 30 \\
	SUMSS & 843 & 45$''$ & 211,050 & 1.25 \\
	NVSS & 1400 & 45$''$ & 1,773,484 & 0.4 \\
	\hline \hline
	\end{tabular}
\label{table:catcomp}
\end{table*}

\section{Creating a Combined Catalogue}
\label{sec:PUMAcat}
During the first incarnation of PUMA, the MWA had been collecting EoR data for months already. The software being used by the Australian MWA EoR pipeline, the Real Time System~\cite[the RTS,][Mitchell et al. 2016 in prep]{Mitchell2008}, requires an input source catalogue to generate a sky model for visibility calibration and source subtraction. Previously, the Molonglo Reference Catalogue~\cite[the MRC,][]{Large1981} was used. Being a shallow and directed survey, the MRC is only complete to $\sim$1~Jy/beam, and reports a single frequency of $408\,$MHz. The MWA observes between 80-$300\,$~MHz, so an assumed spectral index of -0.7 was used to extrapolate the sources to the desired frequencies.

The MWA Commissioning Survey~\cite[MWACS,][]{Hurley-Walker2014a} was undertaken to provide a better model of the sky at these frequencies, during the commissioning phase of the MWA, when the instrument had less elements and thus less sensitivity and poorer resolution. Given that accurate positions were unobtainable with shorter baselines and a new array to calibrate, MWACS was limited in its accuracy, and therefore a prime candidate for testing PUMA.

\begin{figure}
\centering
   \begin{subfigure}{0.45\textwidth}
   \center
   \includegraphics[width=\textwidth]{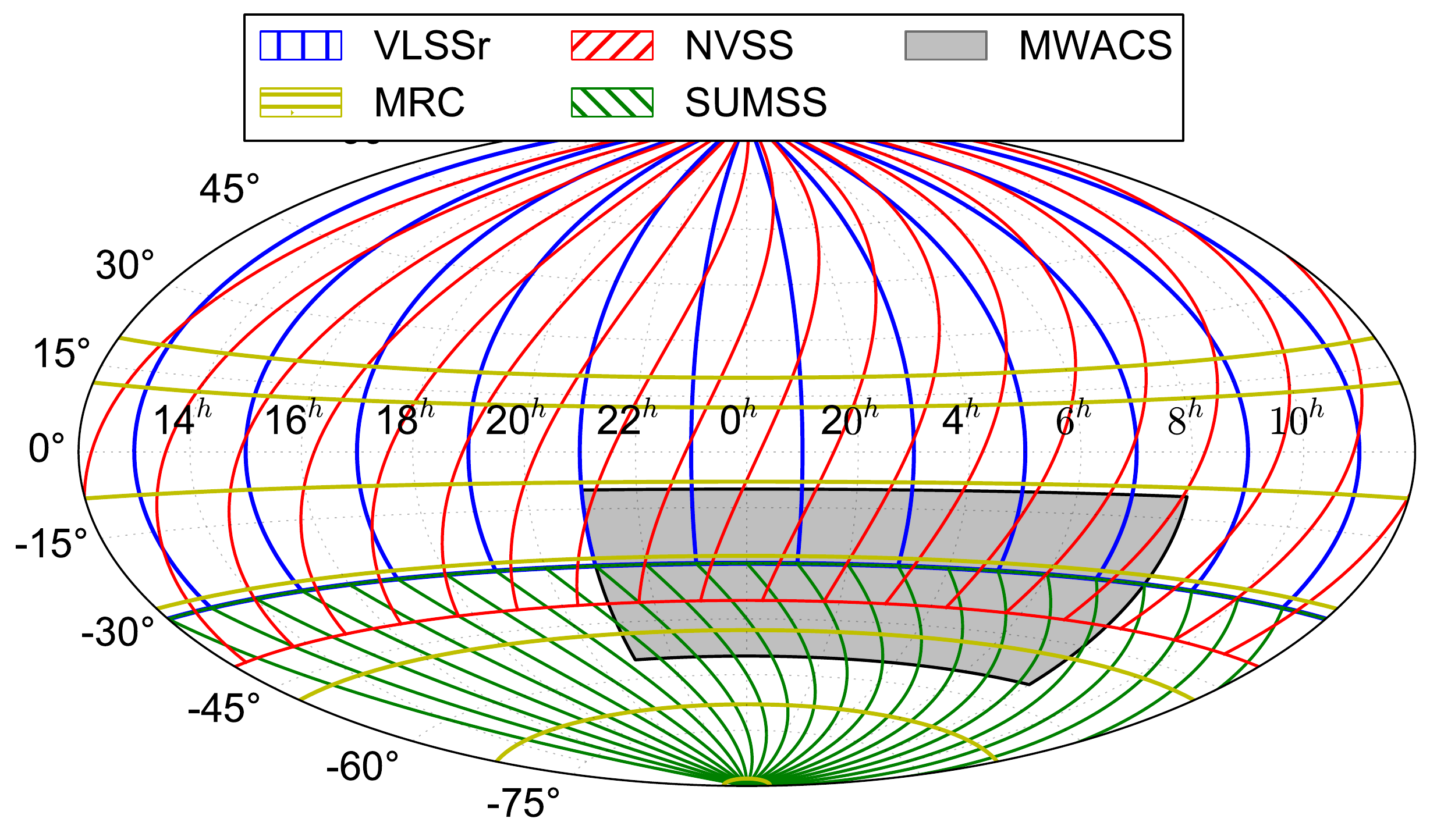}
   \caption{\sffamily{Overall sky coverage}}
   \label{fig:skycoverA}
   \end{subfigure}
   \begin{subfigure}{0.4\textwidth}
   \center
   \includegraphics[width=\textwidth]{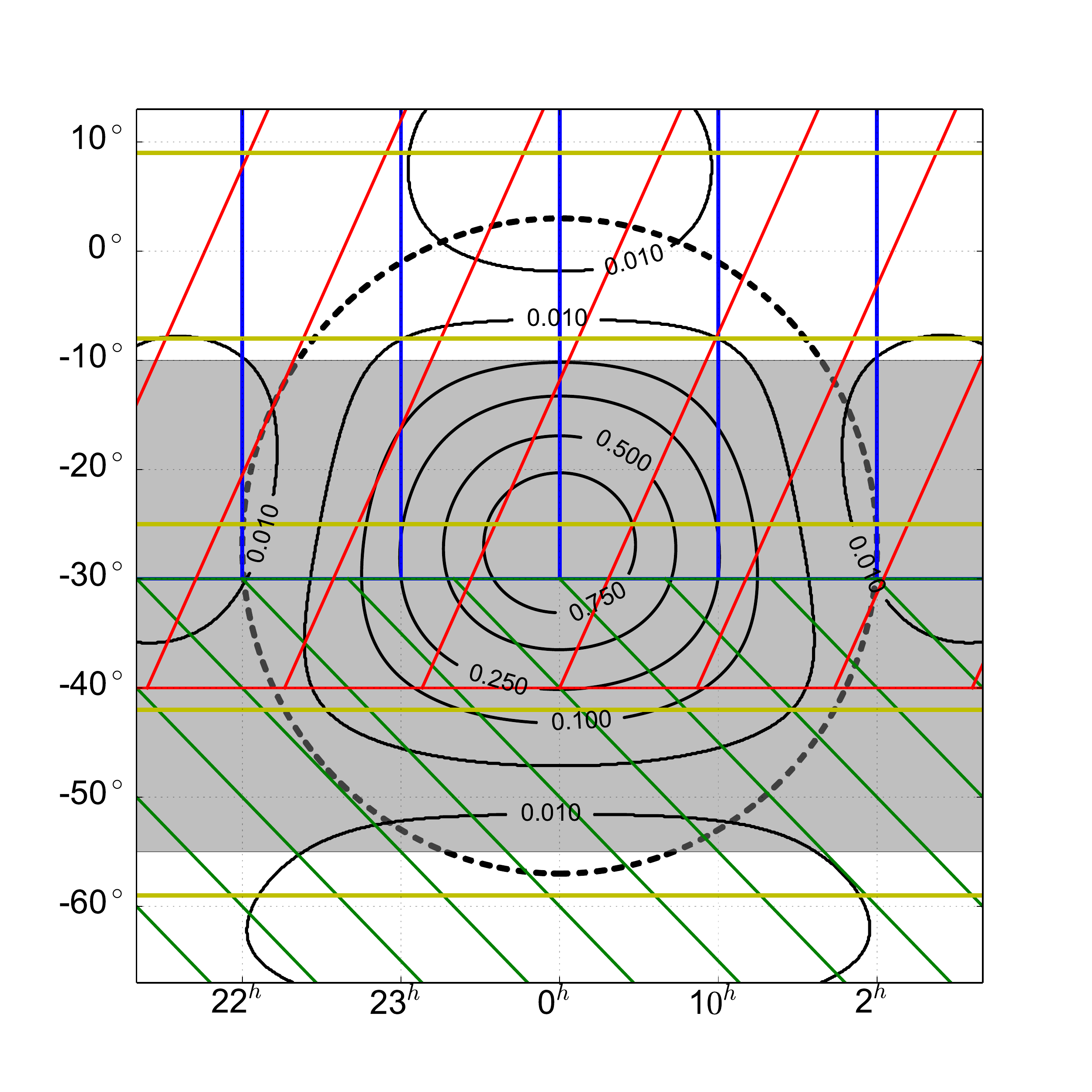}
   \caption{\sffamily{The EoR0 field with example MWA beam}}
   \label{fig:skycoverB}
   \end{subfigure}
\caption{\textsf{The overall sky coverage of each catalogue is shown in~\ref{fig:skycoverA}. Apart from MRC, all catalogues only partially cover the MWACS field, which is emphasized by the zoom in on the EoR0 field shown in~\ref{fig:skycoverB}. A contour plot shows the MWA primary beam at $180\,$MHz, with EoR0 at zenith. The first four grating side lobes are clearly visible outside the dashed circle which represents 2$^h$ from field centre.}}
\label{fig:skycover}
\end{figure}

To best calibrate and remove sources from a target field, a source catalogue must cover the entire field. The first science field targeted for EoR by the MWA was labelled EoR0, centred at 0$^h$, $-$27$^\circ$. Due to the large primary beam of the MWA tile response~\citep[e.g.][]{Tingay2012}, good knowledge of the sky is needed to at least 2$^h$ distance from field centre (see Figure~\ref{fig:skycover}). Further, due to the grating side lobes of the primary beam, power from sources as far as the horizon can enter the visibilities, necessitating a catalogue that covers the entire sky. MWACS unfortunately only extends up to a declination of $\sim-$15$^\circ$ . To get the best combination between depth and coverage at the correct frequencies, it was decided to use MWACS as the base catalogue, and fill any missing sky coverage with MRC as a base.

The following catalogues were used to cross-match: the $74\,$MHz Very Large Array Low Frequency Sky Survey redux~\citep[VLSSr,][]{Lane2012}; the $843\,$MHz Sydney University Molonglo Sky Survey~\citep[SUMSS,][]{Mauch2003}; and the $1.4\,$GHz NRAO VLA Sky Survey~\citep[NVSS,][]{Condon1998}. These surveys were selected due to their frequencies and sky coverage (see Figure~\ref{fig:skycover} and Table~\ref{table:catcomp}). 

\subsection{Running PUMA}
\label{subsec:runPUMA}
PUMA was run first using MWACS as base and matching to all other catalogues, then again using any MRC sources lying outside of the MWACS field as a base. In both cases, the following parameters were used: $P_u = 0.95$; $P_l = 0.8$; $\chi_{red,u}^2 = 10$; $\epsilon_u = 0.1$. These settings were decided upon after investigating matching outcomes; the chosen spectral fitting cut-offs allow for some inherent curvature of the SED, whilst still failing for large deviations from linearity. For an exploration of the effect of these parameters on the PUMA classifications, see Appendix~\ref{app:params}. As the splitting test outlined in \ref{subsec_MulMatch} only covers the most simple cases of repeated catalogue matches, this option was not invoked during this analysis. The order of catalogue position preferred was set as NVSS; SUMSS; MWACS/MRC. VLSSr was not selected as a correcting catalogue as it suffered from ionospheric effects and has already been positionally fit using NVSS \citep[see][for details]{Cohen2007a,Lane2014}. The matching statistics from both runs are shown in Table~\ref{table:finalcat}, showing that overall 98.6\% of sources were accepted by PUMA and automatically matched. Table~\ref{table:catcomp} also shows that some MWACS or MRC sources found no match at all with other catalogues. After investigation, the vast majority of these sources were found to lie near the galactic plane in the southern hemisphere. SUMSS does not cover the sky over $|b| < 10^\circ$; as only SUMSS and MRC extend below $\delta = -40^\circ$,  this accounts for the missing matches. Given the galactic plane lies more than 4$^h$ away at the declinations applicable to the EoR0 field, these sources were ignored.

\begin{table}
\centering
\renewcommand{\arraystretch}{1.1}
\caption{\textsf{The settings used and matching statistics obtained when running PUMA on real data. The number of sources shows the number of base catalogue sources for each case, and the number of matches the instances where a match to at least one catalogue was found.}}
	\begin{tabular}{l l l}
	\hline
	\textbf{Parameter} & \textbf{MWACS base} & \textbf{MRC base} \\
	\hline \hline
	Resolution (arcmins) & 03:00 & 02:48 \\
	Number sources & 14,111 & 9488 \\
	Number matches & 13,995 & 8880 \\
	\hline
	\texttt{accept} & 13,785 & 8691 \\
	\qquad as \texttt{isolated} & \qquad 10,486 & \qquad 6649 \\
	\qquad as \texttt{dominant} & \qquad 1301 & \qquad 904 \\
	\qquad as \texttt{multiple} & \qquad 1998 & \qquad 1138 \\
	\hline
	\texttt{reject} & 20 & 25 \\
	\texttt{eyeball} & 190 & 164 \\

	\hline \hline
	\end{tabular}
\label{table:finalcat}
\end{table}

\subsection{PUMA outcomes}
\label{subsec:PUMAoutcomes}
Of the 22,476 automatically accepted matches, 76\% were classified as \texttt{isolated}, 10\% as \texttt{dominant}, and 14\% as \texttt{multiple}. Figures~\ref{fig:goodisolated}-\ref{fig:eyeball} show example PUMA outcomes for each classification. In each Figure, the two left-hand plots show the positional and spectral information fed into PUMA. If a catalogue included Gaussian fit parameters for sources, these are plotted for instruction only; they were not used by PUMA during the cross-matching process. The plots on the right-hand side then show each cross-match combination of the matched sources, with the pertinent matching probability and spectral fit residuals. 

\subsubsection{SI distribution}
\label{subsubsec:SIdist}
To check for any systematic biases in the differing matching classifications, the SI distributions of each classification were compared in Figure~\ref{fig:SIdist}. The \texttt{isolated} and \texttt{multiple} classifications report near identical distributions, with the \texttt{dominant} resulting in a similar distribution centred at a slightly steeper SI. There are multiple factors that could account for this offset, two contributory factors of which are: the \texttt{dominance} test could consistently choose dominant sources when it should be combining flux densities, thus under-estimating flux densities at high frequencies; the weighted least squares fit is strongly biased to higher flux densities due to the smaller associated errors, and so small changes in flux density at those frequencies greatly affect the fit. To do any accurate modelling of spectra~\citep[e.g.][]{Callingham2015} it is essential to remove confusing sources. This possible steepening is further investigated in \S\ref{subsec:PUMAcomp}.

\subsubsection{Positional Offsets}
\label{subsubsec:posoff}
\begin{figure}
\centering
	\begin{subfigure}{0.9\columnwidth}
	\includegraphics[width=\textwidth]{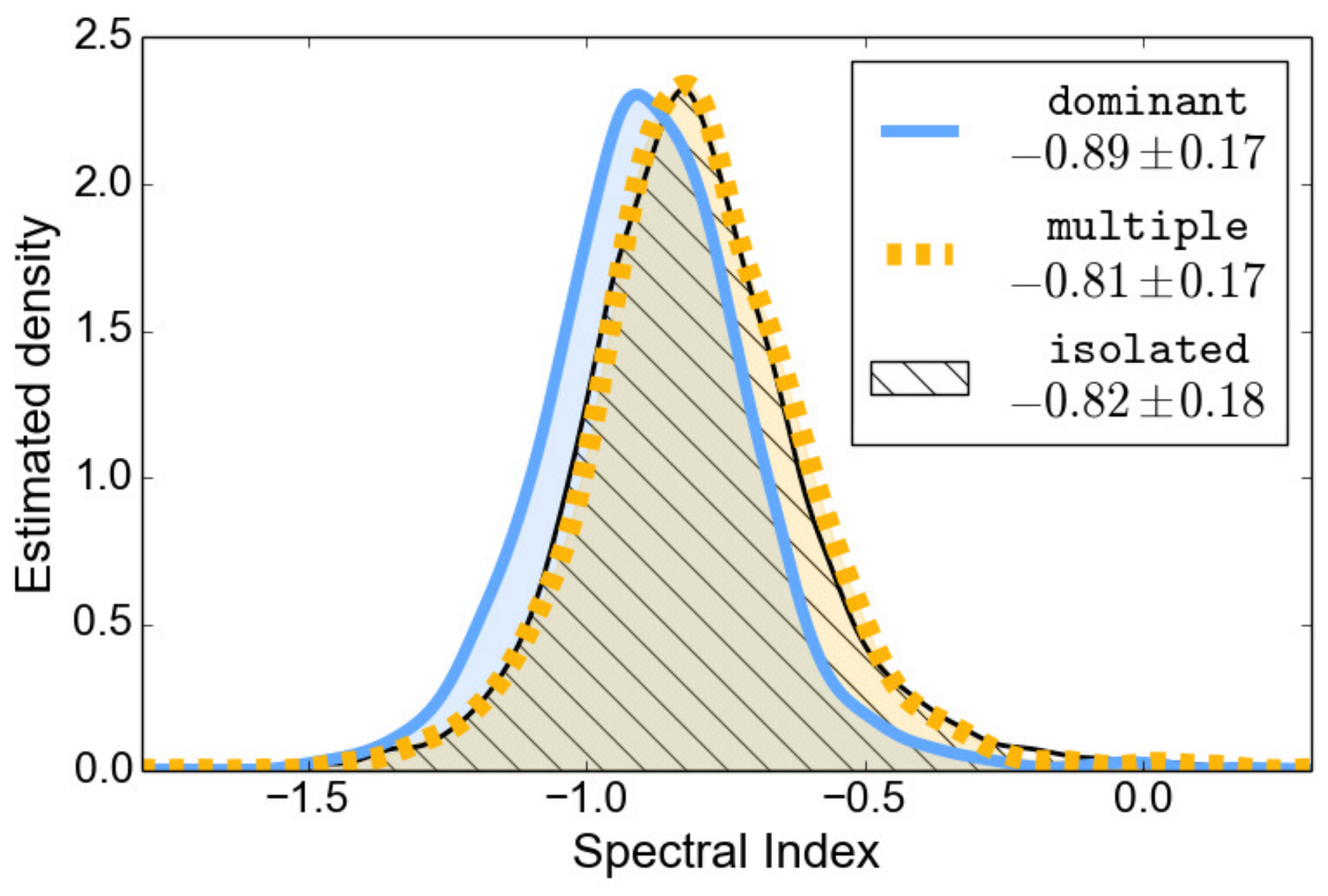}
	\caption{\textsf{SI distribution of all \texttt{accept} sources}}
	\label{fig:SIdist}
	\end{subfigure}
	\begin{subfigure}{0.9\columnwidth}
	\includegraphics[width=\textwidth]{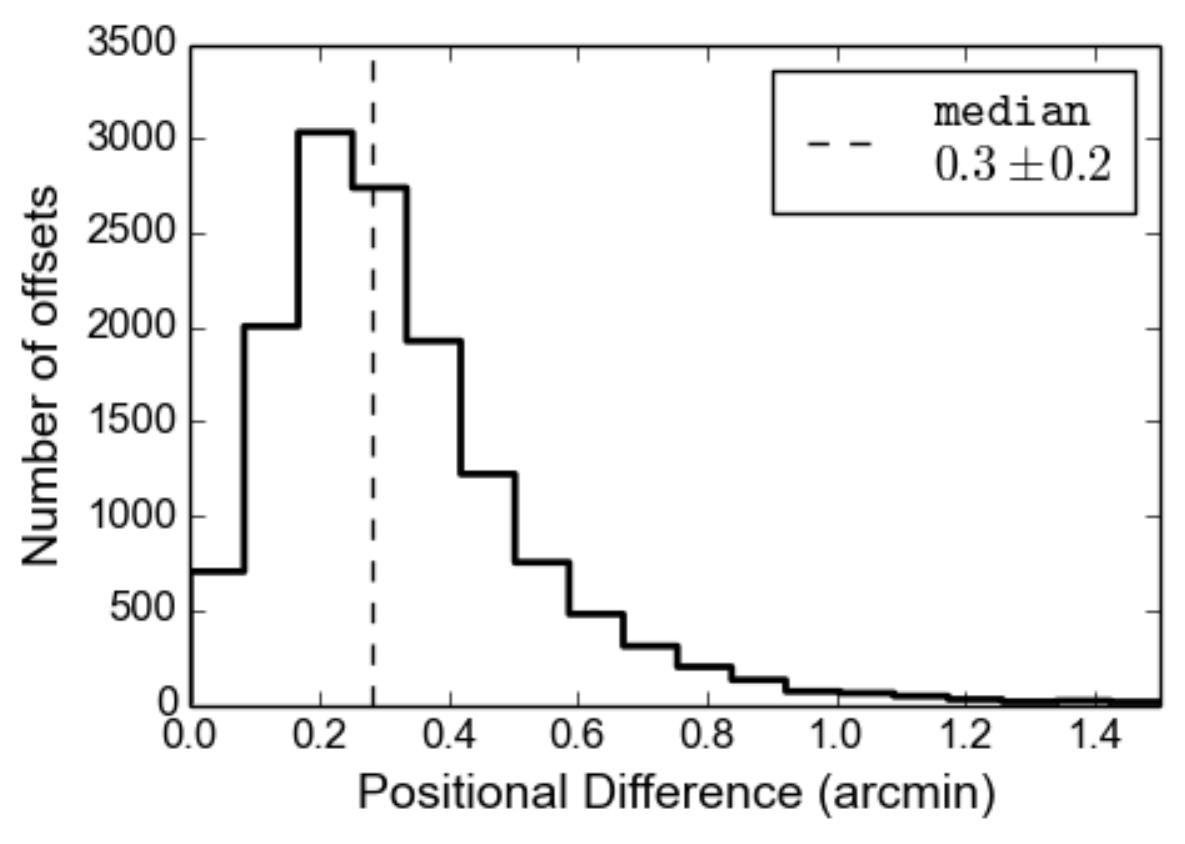}
	\caption{\textsf{Positional offsets of MWACS sources}}
	\label{fig:offhist}
	\end{subfigure}
\caption{\textsf{(a) kernel density estimate of the SI distribution of each PUMA classification. The median and absolute median deviation of each distribution is quoted in the legend. (b) A histogram of the offsets of MWACS sources to either NVSS or SUMSS found by PUMA, including all match types except from \texttt{eyeball} and \texttt{reject}. We find similar positional offset behaviour from NVSS and SUMSS as is described in~\citep{Hurley-Walker2014a}.}}
\end{figure}

\begin{figure*}
\includegraphics[width=1.8\columnwidth]{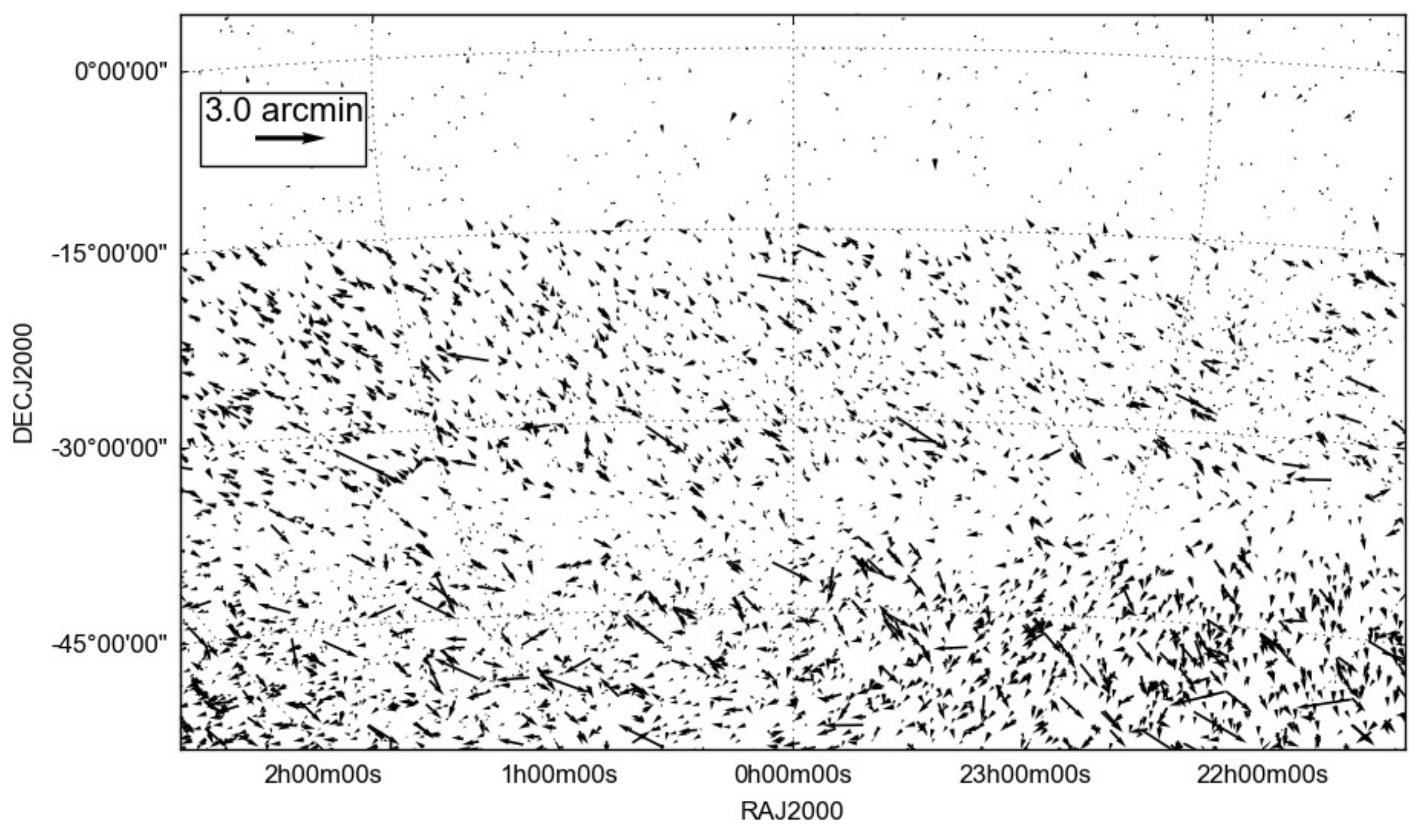}
\caption{\textsf{The positional offset found to either NVSS or SUMSS from either MWACS or MRC is shown. The edge of the MWACS field is clearly seen at $\delta = -15^\circ$. The positional agreement with MRC is excellent, most likely due to MRC only containing bright sources. As explained in~\citealp{Hurley-Walker2014a}, the positional offsets to MWACS vary with RA. The MWACS survey was taken over two declination strips, the effect of which appears to be visible in the plot, with the decrease in offset density at around $\delta = -37^\circ$. There are hints of an overall north-east offset in the upper declination strip; this is further investigated in ~\citep{Carroll2016}. Coherent patches of positional offsets are consistent with a phase gradient introduced by ionospheric effects. As these would vary over a night, the offsets seen here could well be ionospheric.}}
\label{fig:posoff}
\end{figure*}

The positional offsets found when considering only cross-matches with MWACS as a base that were accepted by PUMA are shown in Figure~\ref{fig:offhist}.
The positional offsets for all \texttt{isolated} sources surrounding the EoR0 field are shown in Figure~\ref{fig:posoff}. PUMA classifications are most useful \textbf{in this} instance for quickly flagging out confused cross-matches; the \texttt{isolated} matches can then be used to investigate inherent catalogue properties.

\begin{landscape}
\quad \linebreak
\quad \linebreak
\begin{figure}[h]
\center
\includegraphics[width=1.35\textwidth]{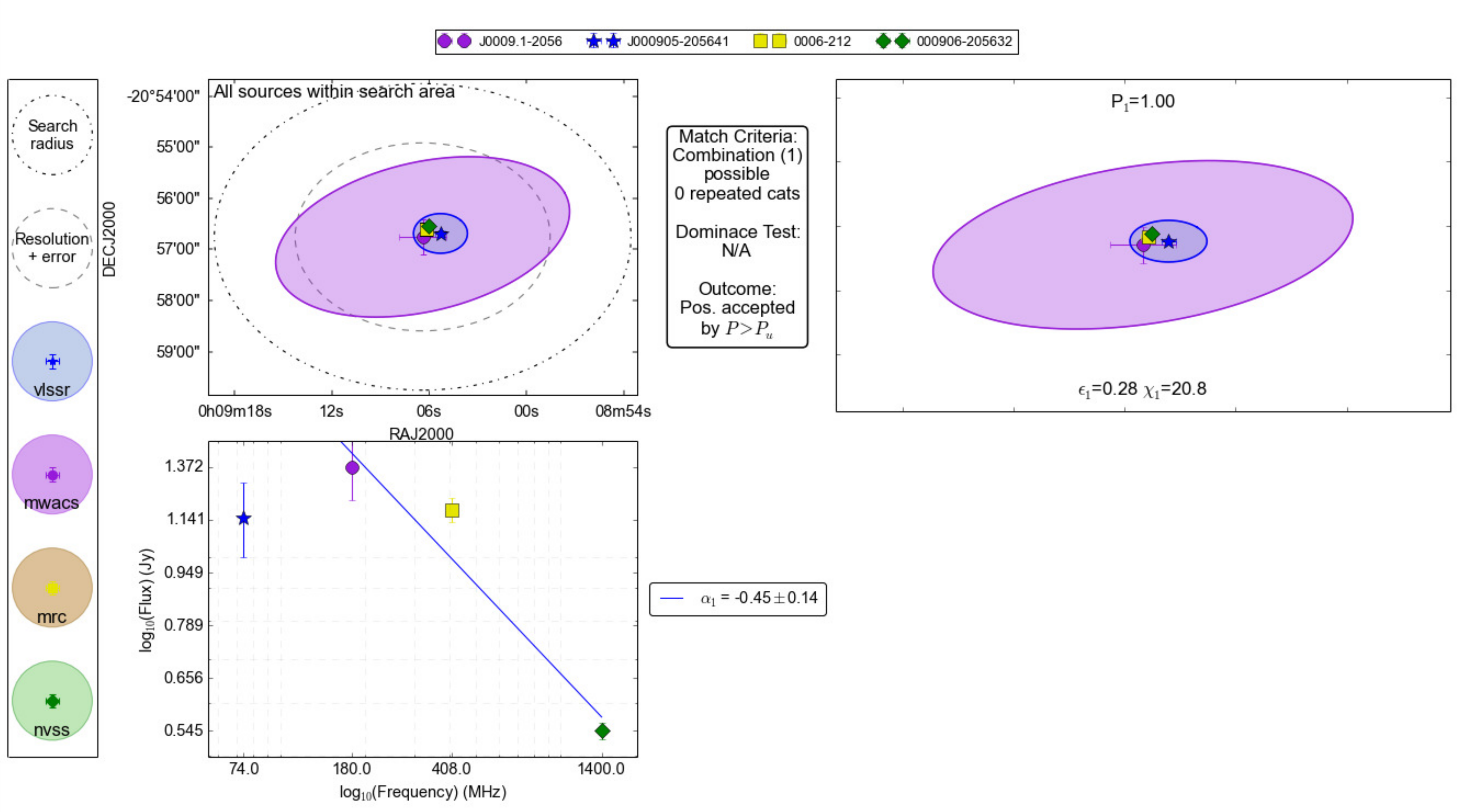}
\caption{\sffamily{An example of an \texttt{accepted isolated} match. As there is only one possible combination of sources, and that combination has $P_1 > P_u$, the cross-match combination is accepted without investigating the SED.}}
\label{fig:goodisolated}
\end{figure}
\end{landscape}

\begin{landscape}
\quad \linebreak
\quad \linebreak
\begin{figure}[h]
\center
\includegraphics[width=1.35\textwidth]{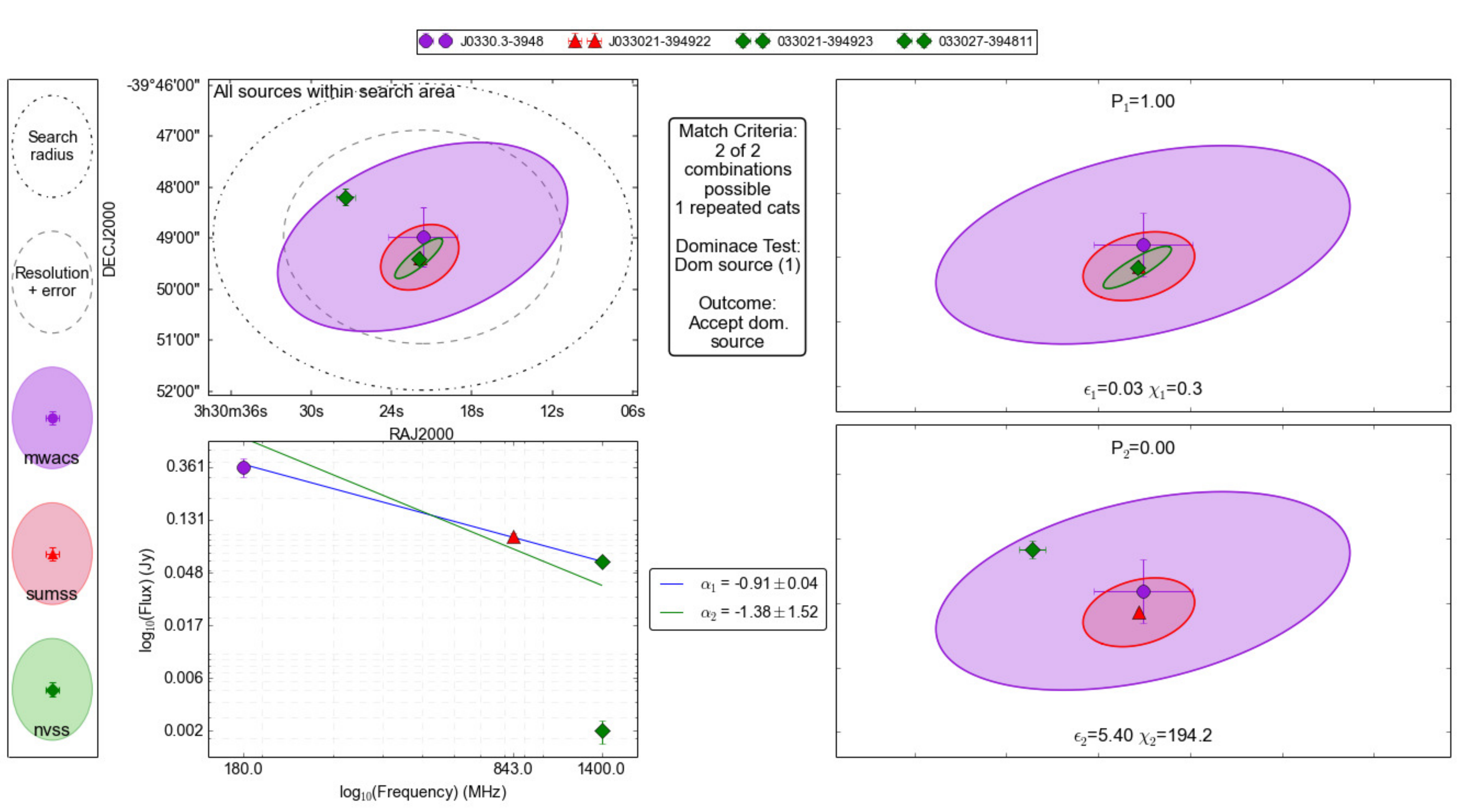}
\caption{\sffamily{An example of an \texttt{accepted dominant} match. There are two NVSS sources well within the resolution of the base MWACS source. Given the positional error on the MWACS source, both cross-match combinations yield high positional probabilities. The SEDs of both cross-match combinations are investigated, and it is found that $P_1 > P_u$, $P_2 < P_l$ as well as cross-match combination 1 having far lower residuals to a power law than cross-match combination 2. This results in match 1 being selected as the correct match.}}
\label{fig:gooddominate}
\end{figure}
\end{landscape}

\begin{landscape}
\quad \linebreak
\quad \linebreak
\begin{figure}[h]
\center
\includegraphics[width=1.35\textwidth]{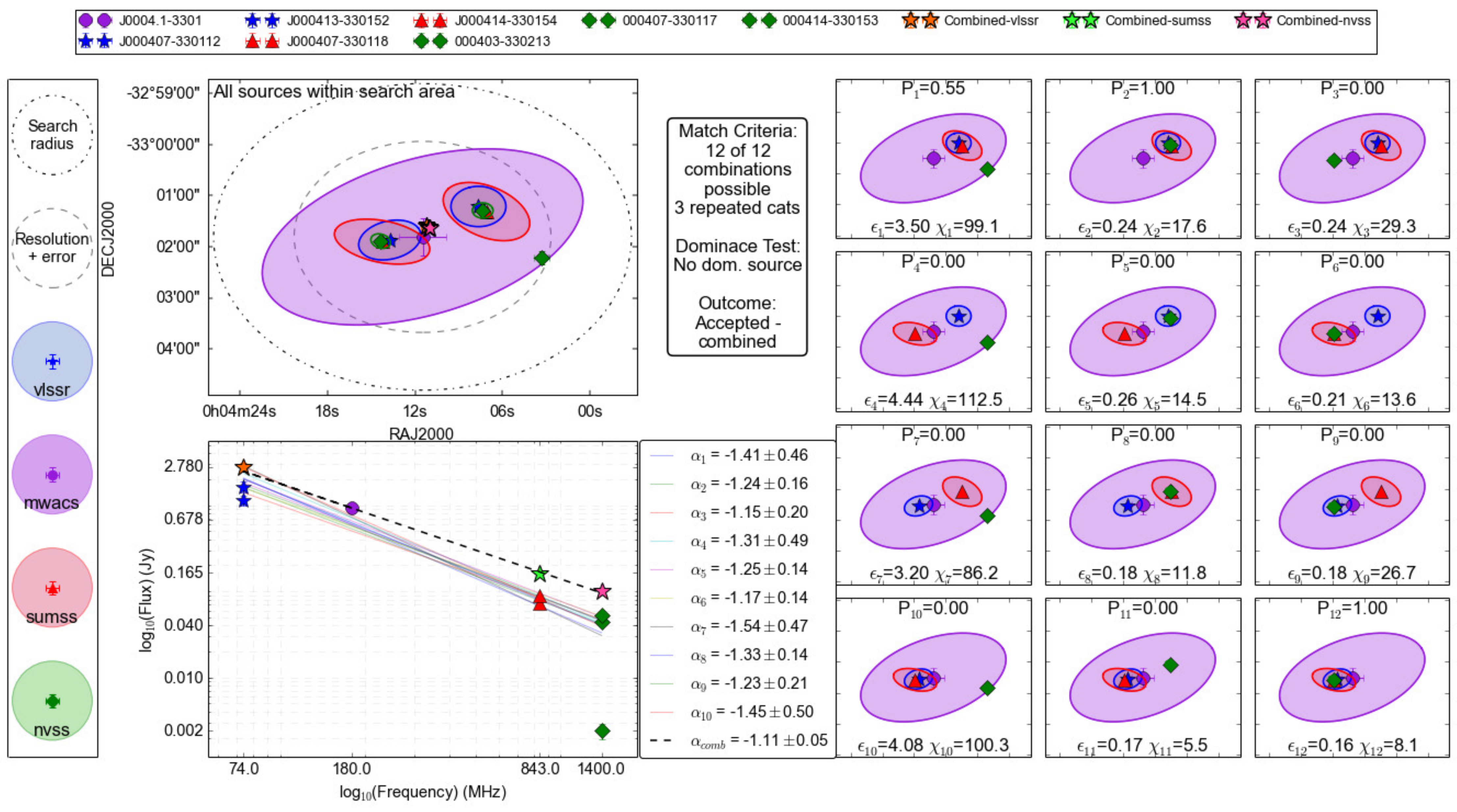}
\caption{\sffamily{An example of an \texttt{accepted multiple} match. In this example both cross-match combinations 2 and 12 yield $P > P_u$ and so there is no \texttt{dominant} match. Instead, all flux densities are combined, and the new SED tested with a power law fit. As the fit is deemed to be good, the source is accepted, and the weighted NVSS position (orange star) is used as the corrected position.}}
\label{fig:goodcombine}
\end{figure}
\end{landscape}

\begin{landscape}
\quad \linebreak
\quad \linebreak
\begin{figure}[h]
\center
\includegraphics[width=1.35\textwidth]{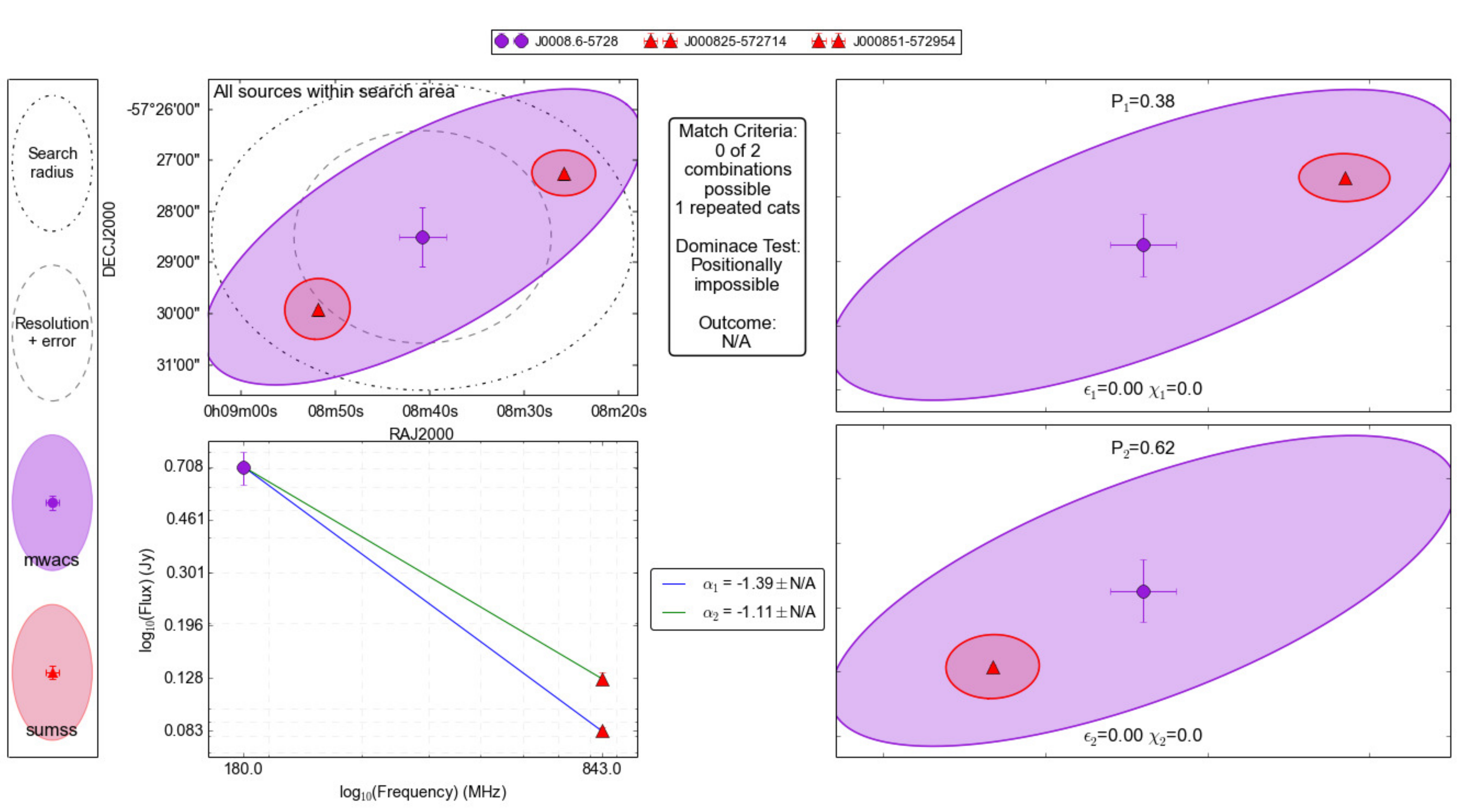}
\caption{\sffamily{An example of an \texttt{reject position} match. Both SUMSS source lie outside of the resolution of the MWACS catalogue plus the positional error of the MWACS source. As $P_{1,2} < P_u$, all cross-match combinations are deemed improbable and are rejected. Further investigation of cross-matches such as these are best diagnosed in conjunction with postage stamp images such as shown in Figure~\ref{fig:extplot}}}
\label{fig:rejposition}
\end{figure}
\end{landscape}

\begin{landscape}
\quad \linebreak
\quad \linebreak
\begin{figure}[h]
\center
\includegraphics[width=1.35\textwidth]{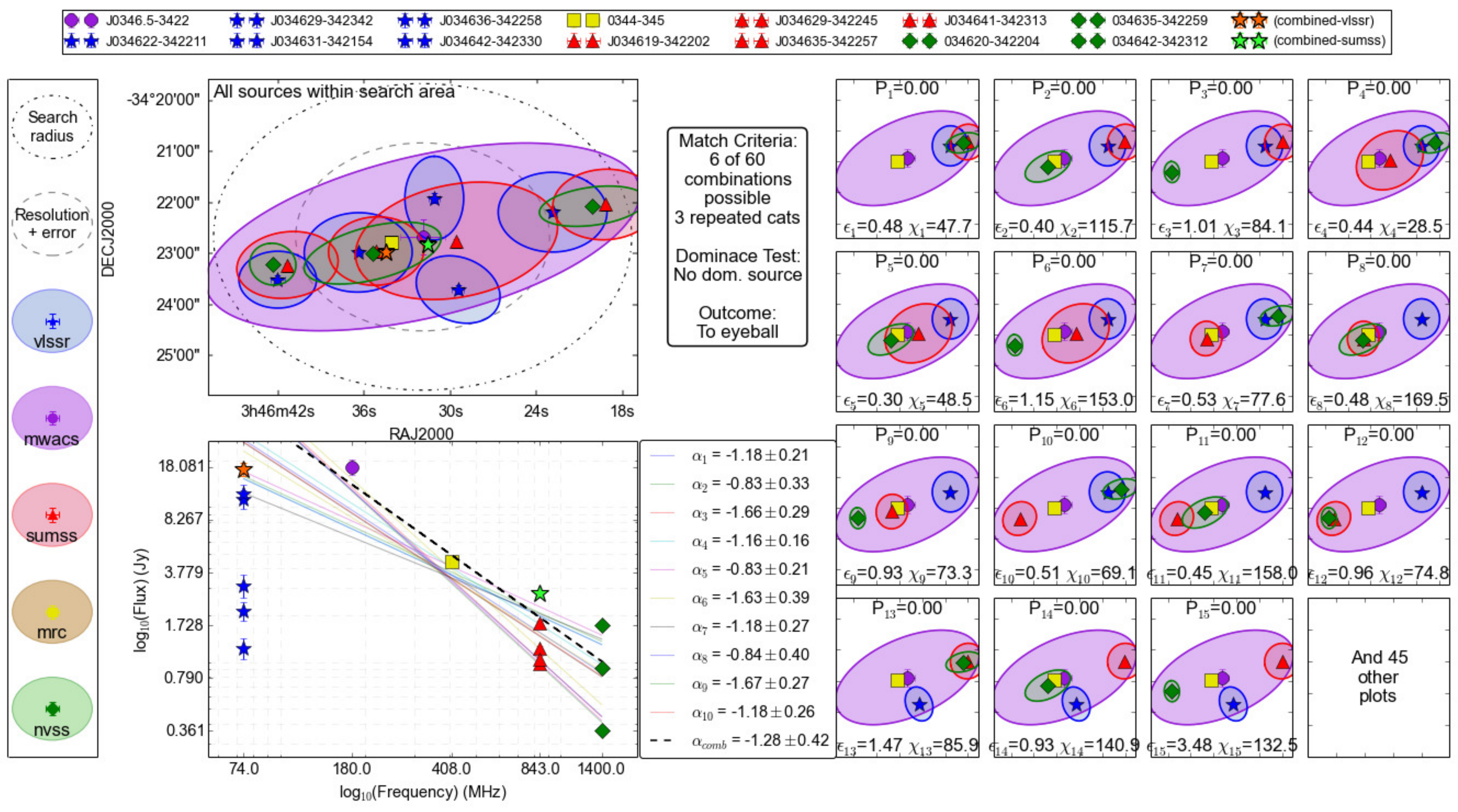}
\caption{\sffamily{An example of an \texttt{eyeball multiple} match. Many cross-match combinations lie outside of the resolution plus error of the base MWACS source, with no \texttt{dominant} combination. A sum of the flux densities of the matched sources that passed the positional criteria yields a poor fit to a power law, and so the MWACS source is not accepted and labelled to \texttt{eyeball}. Again, further investigation of cross-matches such as these are best diagnosed in conjunction with postage stamp images such as shown in Figure~\ref{fig:extplot}.}}
\label{fig:eyeball}
\end{figure}
\end{landscape}

\subsection{\texttt{eyeball} and \texttt{reject} sources}
\label{subsec:extended}
\begin{figure*}
\centering
\includegraphics[width=1.7\columnwidth]{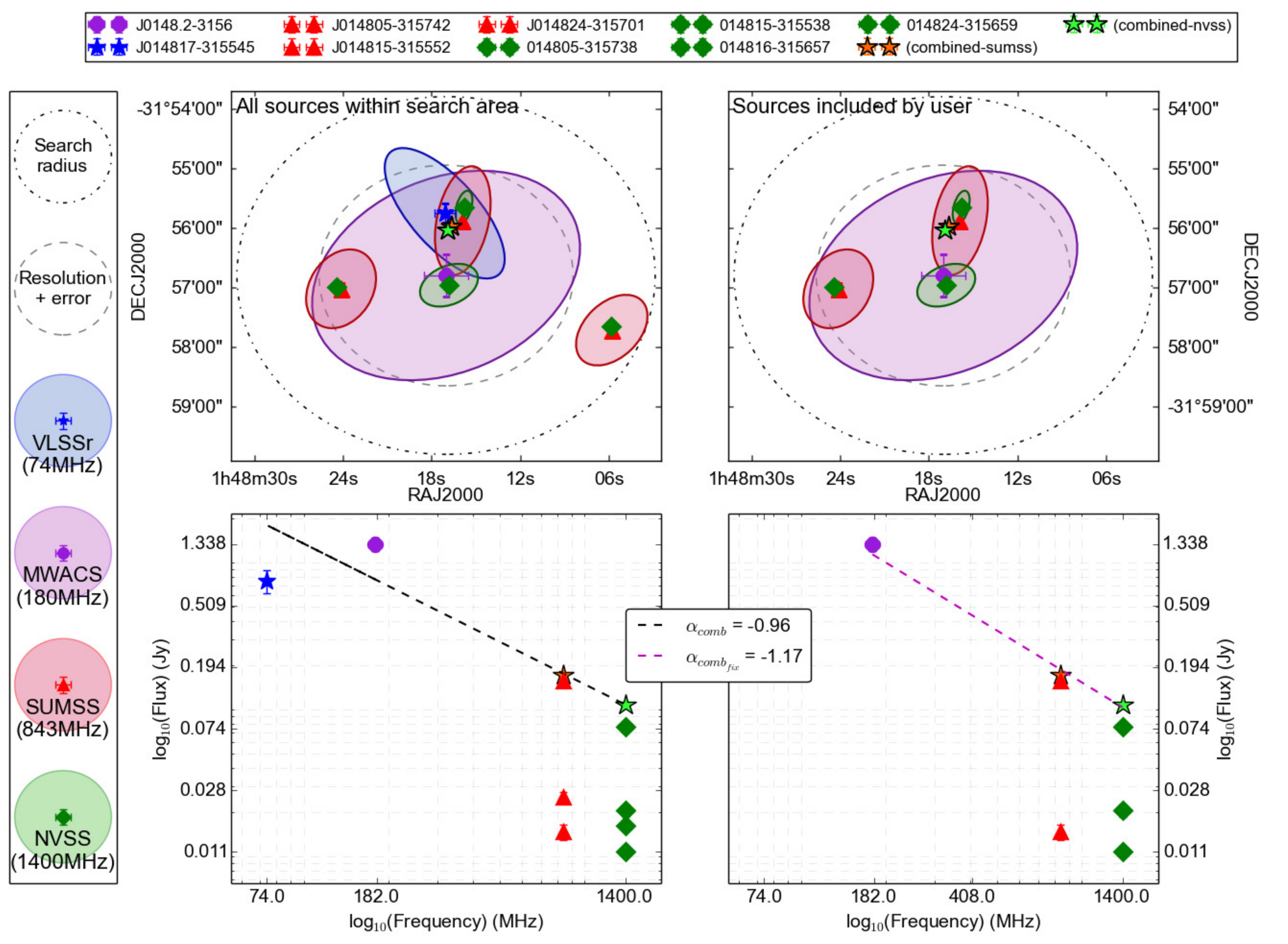}
\includegraphics[width=1.7\columnwidth]{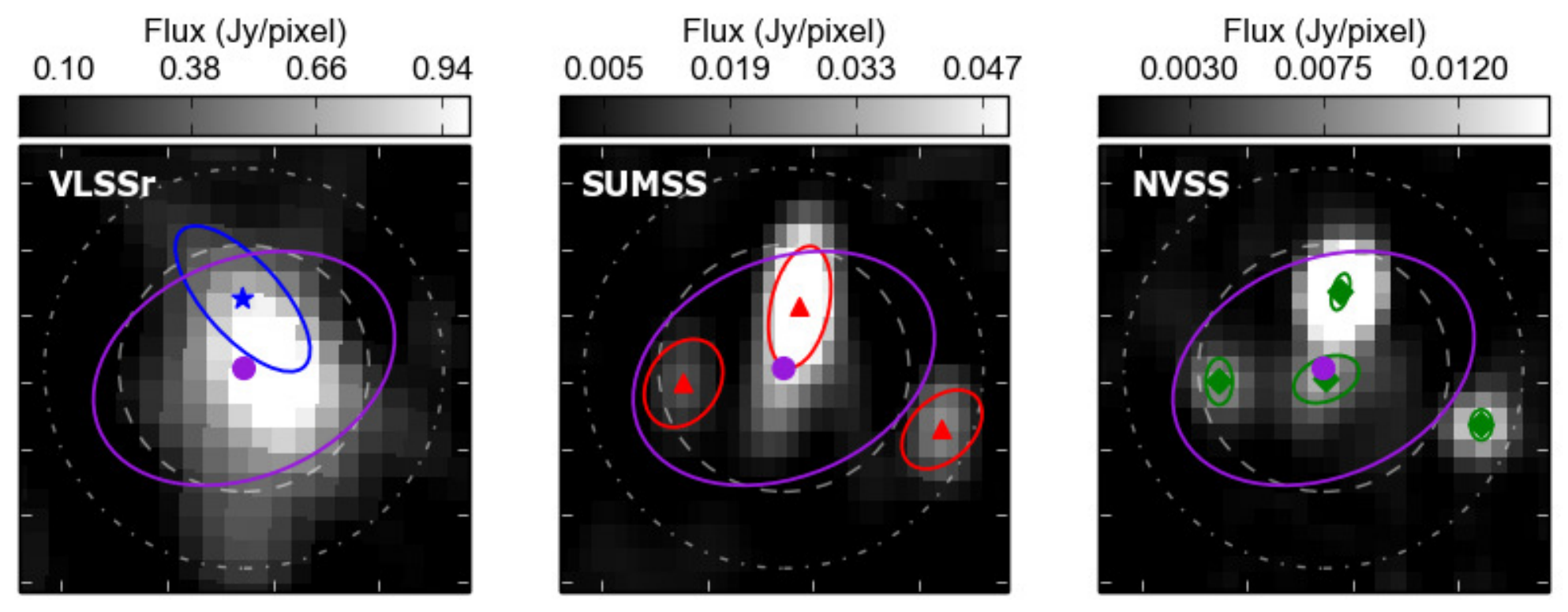}
\caption{\textsf{An example of the matching process for extended sources is shown. The two upper left panels show all information given to PUMA. The bottom three panels show postage stamp images of the three matched catalogues, with the reported catalogues over-plotted, along with the MWACS source. In this case, the source reported in the VLSSr catalogue does not realistically match the VLSSr image. This artificially creates a curved SED, which causes PUMA to label this match an \texttt{eyeball}. Given the doubt cast on the VLSSr source, it is ignored in the cross-match, and the SUMSS and NVSS sources that seem positionally reasonable are combined and matched. This gives a realistic positional match as well as spectra.
}}
\label{fig:extplot}
\end{figure*}

399 sources were not automatically catalogued by PUMA. As this matching process was run with the EoR0 field in mind, it was decided to include sources within 2$^h$ of the EoR0 field centre, leaving 74 sources to investigate. For each flagged source, postage stamp images of available relevant catalogues were obtained and plotted. These were used in conjunction with catalogue information to make an informed decision on a cross-match. An example of this process is shown in Figure~\ref{fig:extplot}. The MWA data simulated in \S\ref{sec:PSanaly} was taken when the MWA had greater resolution ($\sim2.3'$) than when MWACS was created. For this reason, any MWACS source that was matched to multiple components that were separated by an angular distance that would be resolved at $2.3'$ were split into multiple catalogue entries in a bid to reduce residuals after source subtraction when using real data.

\section{Testing PUMA}
\label{sec:testPUMA}

To test the fidelity of PUMA, it is necessary to know the true sky. The known positional and SI distributions can then be used to compare to the positional corrections and SI distribution derived by PUMA. In this section, a point source sky model is created using the NVSS catalogue. It is used to create simulated sky images as seen by the 5 telescopes for the catalogues matched in \S\ref{sec:PUMAcat}. Source-finding is then applied to these images to create 5 mock catalogues on which to test PUMA. The two criteria used to test PUMA is the recovery of accurate positions and SI.

\subsection{Mock Catalogues}
\label{subsec:mock_catalogues}

\subsubsection{NVSS sky model}
\label{subsubsec:nvss_sky_model}
NVSS catalogued positions and flux densities were selected from a fiducial lune of sky, bounded by $0.0^\circ \leq RA \leq 30.0^\circ$ and $-40.0^\circ \leq \delta \leq -30.0^\circ$. This patch of sky was chosen as it lies in the overlap region between SUMSS and NVSS, and to generate enough sources to have statistical significance when testing PUMA. To get a realistic SI for each point source, a simple positional cross-match was performed to SUMSS, with a cut-off of half the beam width of the NVSS survey ($22.5''$). Any NVSS source without a match was assigned a random SI, drawn from a gaussian distribution with $\mu = -0.8, \sigma = 0.2$, to reflect the SI distributions seen in Figure~\ref{fig:SIdist}.

\subsubsection{Simulated sky images}
\label{subsubsec:sim-sky-images}
\begin{figure}
\centering
\includegraphics[width=\columnwidth]{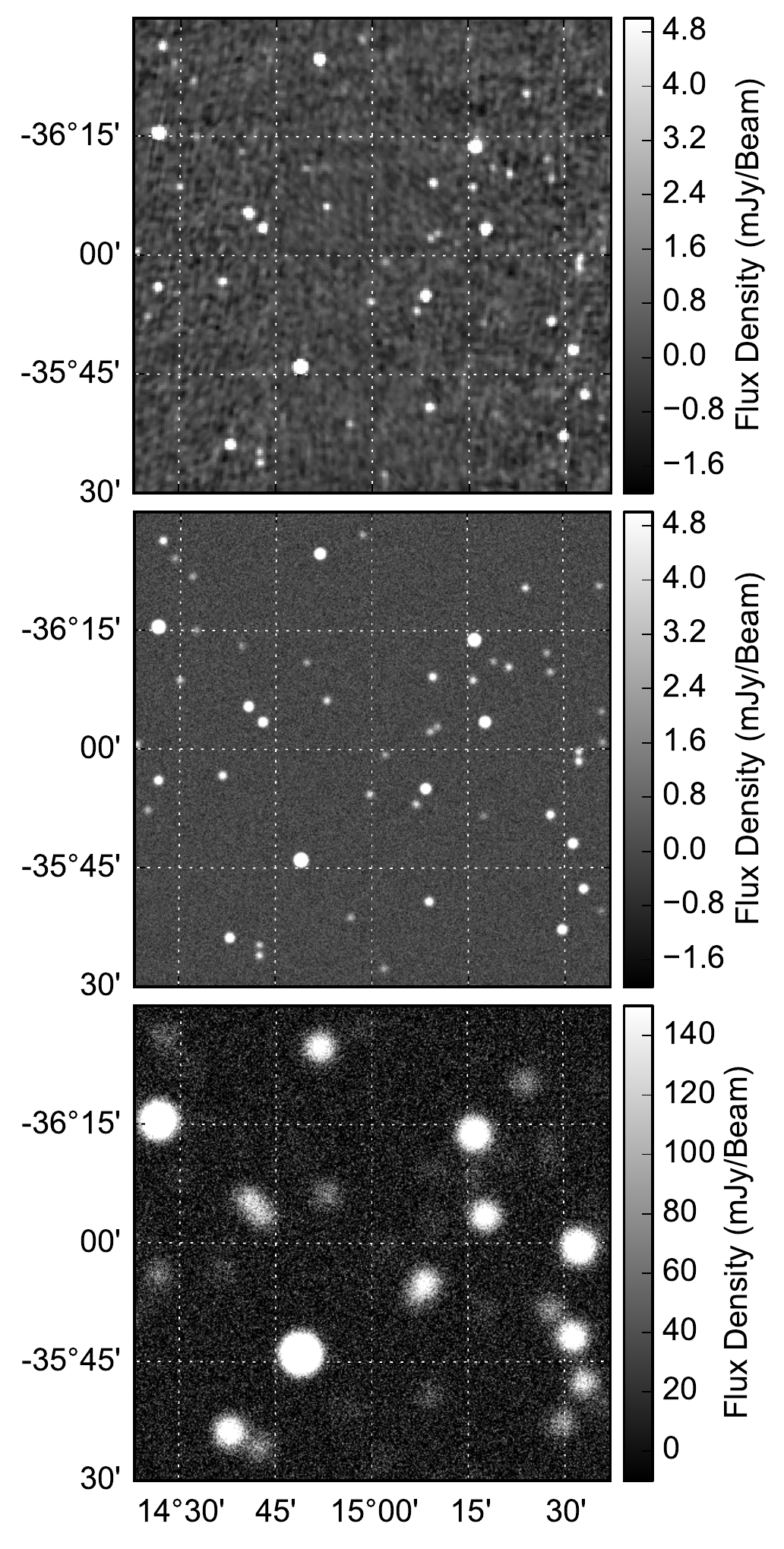}
\caption{\textsf{A comparison of a real NVSS postage stamp image (upper panel) and a simulated NVSS postage stamp (central panel), created as described in \S\ref{subsubsec:sim-sky-images}. The same area of sky is also shown as simulated to mimic an MWACS postage stamp (lower panel).
}}
\label{fig:sim-sky-comp}
\end{figure}

The selected NVSS sources were used to populate a point source sky array model at the frequencies of the 5 catalogues listed in Table~\ref{table:catcomp}, extrapolating the flux densities using the assigned SI. The \texttt{astropy.convolution}\footnote{\texttt{astropy} convolution documentation - \url{http://docs.astropy.org/en/stable/convolution/}} python library was then used to convolve the sky array with a Gaussian kernel, set to have a full width half maximum value equal to the beam width listed in Table~\ref{table:catcomp}. In doing so, the restoring phase of a \texttt{CLEAN}~\citep{Hogbom1974} image process is mimicked\footnote{Normally the projection of the synthesized beam is applied to the restoring Gaussian, however for simplicity a symmetric Gaussian kernel is used.}; this process is typically applied to interferometric radio data to remove the instrument response from the image, with the restored image then used for source finding.  Gaussian noise was added to the image based on reported image $rms$ from the literature of each catalogue (see Table~\ref{table:catcomp}), to make the source finding as realistic as possible. Figure~\ref{fig:sim-sky-comp} shows a comparison the simulated NVSS to sky to the real data, as well as the simulated MWACS image.

\subsubsection{Source Finding and SI derivation}
\label{subsubsec:sourcefind}
Source finding was performed using the \texttt{PyBDSM} package\footnote{\texttt{PyBDSM} documentation - \url{http://www.astron.nl/citt/pybdsm/}}, which is designed to perform source finding on radio interferometric images, and is capable of fitting Gaussians, Shapelets, and Wavelets to an image. In this analysis only Gaussians were fit to the mock catalogues. To derive the SI of sources found for the mock MWACS catalogue, 5 extra sky images were made at 74, 180, 408, 843, and 1400$,$MHz, without any noise. The Gaussian fit parameters found by \texttt{PyBDSM} for the mock MWACS catalogue were then used to measure the flux density for each source in the noiseless images, by summing the pixels that were bound by the Gaussian fit. These measured noiseless flux densities were then fit using least squares to calculate an expected SI. To ensure that flux densities used to estimate the SI and those used to test PUMA in \S\ref{subsec:PUMAcomp} are consistent, the method of generating a noiseless image and directly summing to measure a flux density was applied to all the mock catalogues; all positional information and errors derived by \texttt{PyBDSM} were retained.

\subsection{PUMA comparison}
\label{subsec:PUMAcomp}
\begin{figure*}
\centering
\includegraphics[width=1.8\columnwidth]{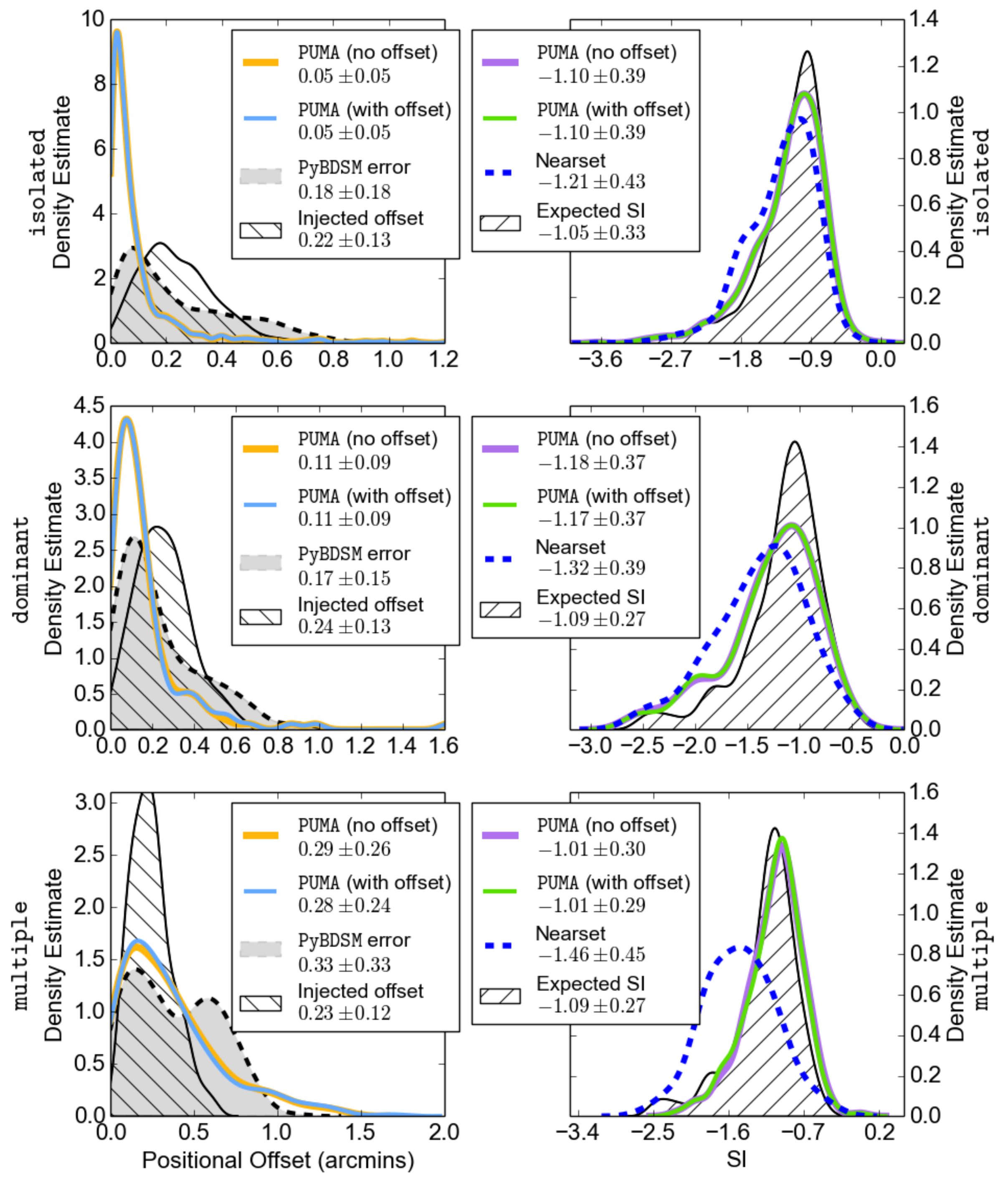}
\caption{\textsf{The positional corrections (left column) and SI distributions (right column) derived by PUMA when matching mock catalogues, split in to \texttt{isolated}, \texttt{dominant}, and \texttt{multiple} (top, middle, and bottom rows, respectively). For every distribution, the median and median absolute deviation is quoted in the legend. In the left hand column, the PUMA positional corrections found when using the \texttt{PyBDSM} mock MWACS positions (no offset) and with positional offsets added (with offset) are shown. The added positional offsets (injected offsets), as well as the \texttt{PyBDSM} reported errors are also plotted. In the right hand column, the PUMA SI distributions are again shown for both the \texttt{PyBDSM} MWACS and the perturbed positions, compared to expected SI distribution as derived in \S\ref{subsubsec:sourcefind}, by finding the flux density from noiseless mock MWACS images. An SI distribution is also shown by performing a nearest neighbour match to the \texttt{PyBDSM} MWACS positions to within $90''$. The PUMA classifications were taken from the match with the original \texttt{PyBDSM} MWACS positions; only matches which were accepted by both PUMA runs (offset and no offset) are plotted for a direct comparison.
}}
\label{fig:mock-cat-outcomes}
\end{figure*}

PUMA was run using the mock MWACS catalogue as a base in the same way as described in \S\ref{subsec:runPUMA}. The matching outcomes are summarised in Table~\ref{table:mockcatcomp}. The source positions and SI found in \S\ref{subsubsec:sourcefind} for the mock MWACS sources were taken to be the `true' source characteristics on which to compare the outputs of PUMA to. To test the robustness of the positional offset recovery, PUMA was run a second time, after random positional offsets were added to the mock MWACS catalogue. The derived positional offsets from the \texttt{PyBDSM} positions and calculated SI are shown for each PUMA classification in Figure~\ref{fig:mock-cat-outcomes}. For comparison, the derived SI from running a simple nearest-neighbour cross-match within $90''$, approximating the FMHW of the mock MWACS beam, is shown. This highlights the power of combining high resolution catalogue data; PUMA reliably retrieves the correct SI for \texttt{mulitple} matches, whereas a simple nearest-neighbour match consistently retrieves a steeper SI. Figure~\ref{fig:mock-cat-outcomes} shows that PUMA behaves the same in the presence of unaccounted positional errors, on top of those quoted by \texttt{PyBDSM}, given the same positional and SI distributions retrieved in both runs. The positional offsets found for \texttt{isolated} sources are small, and are smaller than the \texttt{PYBDSM} errors. Given this, in conjunction with the coherent ionospheric offsets found in the MWACS catalogue, as well as the coherent offsets found by PUMA in~\citet{Carroll2016}, it is clear that for \texttt{isolated} sources, the positional corrections are indeed improving the positional accuracy of the source, whilst reliably reported the correct SI.

If \texttt{dominant} cases are purely discriminating chance alignments of physically unrelated sources, we might expect the positional offsets derived for \texttt{isolated} and \texttt{dominant} to be the same. The distributions are indeed similar, however the \texttt{dominant} distribution shows a median offset of around double that of the \texttt{isolated} cases. This is likely due to the confusing source(s) contributing some flux density to the lower resolution catalogue sources; this blending skews the positional of the base catalogue source. The \texttt{dominant} distribution is closer to the \texttt{isolated} distribution than the \texttt{multiple} distribution however, which lends confidence that the \texttt{dominant} class should be kept separate to the \texttt{multiple}. As seen in Figure~\ref{fig:SIdist}, a slight steepening of the SI distribution is shown for \texttt{dominant} matches. This is also likely due to some blending of sources, which would manifest as an over-estimation of the lower resolution (usually lower frequency) flux density, naturally causing a steepening in the SI. However, as the dominant component should be driving the spectral behaviour of the source at the base catalogue frequencies, this reported SI should realistically describe the behaviour of the source as seen by the base catalogue.

Figure~\ref{fig:mock-cat-outcomes} shows that the positional offsets found for \texttt{multiple} matches are comparatively large. As defined in \S\ref{subsec_MulMatch}, the positional offsets here are derived from the flux density weighted centre of the combined higher resolution catalogue sources: these offsets can then be dominated by the differences in morphology at differing frequencies. As such these positional corrections may not actually be improvements at the frequency of the base catalogue. The method clearly works when estimating the SI however, so it remains up to the user as to which position is appropriate for their desired science case.

\begin{table}
\centering
\renewcommand{\arraystretch}{1.2}
\caption{\textsf{The matching classifications found by PUMA when matching the mock catalogues (No offset), along with the case where positional errors were introduced into the mock MWACS catalogue (With offset).}}
	\begin{tabular}{l l l}
	\hline
	\textbf{Matching class} & \textbf{No offset} & \textbf{With offset} \\
	\hline \hline
	\texttt{accept} & 2129 & 2119 \\
	\qquad as \texttt{isolated} & \qquad 1781 & \qquad 1771 \\
	\qquad as \texttt{dominant} & \qquad 129 & \qquad 127 \\
	\qquad as \texttt{multiple} & \qquad 219 & \qquad 221 \\
	\hline
	\texttt{reject} & 20 & 31 \\
	\texttt{eyeball} & 32 & 32 \\

	\hline \hline
	\end{tabular}
\label{table:mockcatcomp}
\end{table}

\section{Simulations of Foreground Removal Towards an EoR Detection}
\label{sec:PSanaly}

The red-shifted 21cm Hydrogen emission line (HI line) is a tracer of neutral hydrogen (HI) and as such can be used to measure the effect of ionising radiation on HI during the EoR, and can be used to constrain the timing of the epoch as well as the nature of the first ionising sources. As stated in \S\ref{sec:intro}, extra-galactic radio sources emit at the same frequencies and must either be removed or avoided to make a detection. The MWA EoR pipelines have opted for direct foreground removal in the case of point sources~\citep[][]{Jacobs2016}. In this section we create simulations to investigate how subtracting point sources with inaccurate astrometric information will effect a measurement of the EoR signal.

Current generation low-frequency radio arrays are not sensitive enough to directly image the HI line from the EoR. Instead there is a focus on making a statistical measurement through a power spectrum (PS) analysis of the emission of large sky areas. The final measurement will be a 3D spherically averaged\footnote{As this spherical averaging occurs over $k$-modes derived from frequency measurements, it has to be limited to frequency ranges which correspond to $\Delta z \simeq 0.5$. This limits the effect of cosmic evolution, during which time significant changes to the distribution of HI can occur due to on-going ionisation, fundamentally changing the derived PS.} to 1D PS, but the 2D PS serves as a useful instrument and data diagnostic tool. The 2D PS plots variations in measured power across the frequency response of the instrument against its angular response across the sky, and as such different instrumental effects as well as astrophysical signals should inhabit different parts of the 2D PS~\citep[e.g.][]{Morales2012}. Of particular interest in the literature is the `wedge'
\citep[see][]{Datta2010,Parsons2012a, Vedantham2012, Hazelton2013}, a region in which the emission of point sources is expected to be confined~\citep[although some power from point sources should exist outside of this wedge i.e.][]{Thyagarajan2013}. The area most devoid of foreground contamination is known as the `window', and is targeted for an EoR detection.

A PS estimation of the EoR requires removal or avoidance of bright foregrounds including point sources at the same frequencies as the red-shifted 21cm line. If implemented with sufficient accuracy, direct point source removal promises to open up the largest measurement space in the 2D PS~\citep{Pober2014}. Trott et al. have studied the effect of point source removal in the presence of inaccuracies in the data set itself~\citep{Trott2012}. Given the need for accuracy outlined here, this analysis investigates the effect of positional inaccuracies in the calibration and source removal catalogues on the 2D PS. The focus here is to probe where power is removed from the 2D PS, and how that could effect any possible EoR detection. The drawbacks to testing on real data are numerous however: a real instrument can introduce random and systematic errors; the ionosphere is a constantly changing source of error; the true sky brightness distribution is actually unknown, and consists of point like and diffuse emission with varying spectral behaviours. The latter is the biggest obstacle for this experiment, so simulated visibilities were created using OSKAR\footnote{\url{http://oskar.oerc.ox.ac.uk/}}~\citep{Mort2010}. In this manner, the input data could be completely understood and so the effects of positional inaccuracies be isolated.

To achieve this, two source catalogues were created from the PUMA outputs; one using the original MWACS position\footnote{In the case where an MWACS source was split into multiple components (\S\ref{subsec:extended}), the same position was used in both catalogues.}, and a second with the updated PUMA position. We refer to these as the MWACS source list and the PUMA source list, respectively. Both source lists were given exactly the same spectral information. The Australian MWA EoR pipeline was used: all calibration and source subtraction was run using the Real Time System~\citep[RTS,][]{Mitchell2008}, and all 2D PS were made using the Cosmological HI Power Spectrum Estimator~\citep[CHIPS,][]{Trott2016}. The RTS utilises clustered source calibration~\citep{Kazemi2013}, and can either peel (individually calibrate and subtract a source) or subtract as many sources as required.

In~\S\ref{subsec:MWAdata} we outline the MWA observations that were the basis of our simulations; simulating real observations allowed  comparison with data as shown in Figure~\ref{fig:OSKvsMWA}. In~\S\ref{subsec:OSKAR} we detail the OSKAR simulations used to generate MWA-like data. In \S\ref{subsec:results} we present the results of our analysis and the effects of positionally inaccurate source subtraction on the 2D PS.

\subsection{The MWA}
\label{subsec:MWAdata}
The MWA telescope consists of 128 elements, each of which is made of 16 cross-dipole antennas in a $4\times4$ grid. The dipoles in each of these `tiles' are electronically beam-formed, forming a quantized set of primary beam pointings. The signal path of the MWA extracts 30.72 MHz of bandwidth, which it splits into 768 fine channels of 40 kHz in a Polyphase Filterbank (PFB). These data are combined and averaged to 0.5 seconds in the correlator.~\citep[see][for further details]{Ord2015}.

The MWA is located on the Murchison Radio Observatory site, which is extremely radio-quiet~\citep{Offringa2015,Allison2015}; even so some radio interference remains so all data was flagged using \texttt{COTTER}~\citep[see][]{Offringa2015}. Due to the bandpass imparted by the PFB, 5 out of every 32 fine channels are also flagged.

A fiducial night of data was selected to test the reduction pipelines employed by the MWA EoR analysis team \citep[see][for details]{Jacobs2016}. A subset of 1 hour of these observations were selected for testing here, chosen for balance between integration time and processing costs. The EoR0 field stays within one pointing of the MWA for approximately half an hour, giving two pointings (and therefore two beam patterns) in the dataset. The zenith pointing (e.g. Figure~\ref{fig:skycoverB}) and one pointing 6.8$^\circ$ off zenith were chosen, as the MWA beam is best understood at zenith~\citep{Neben2015}. 

\subsection{OSKAR simulations}
\label{subsec:OSKAR}
OSKAR was primarily created to simulate data from SKA-like interferometric arrays. Accurately simulating visibilities is a computationally expensive endeavour, due to multiple Fourier Transforms and gridding steps. OSKAR deals with this by assigning each point source in the sky model to a thread on a GPU, greatly speeding up the process. Importantly, OSKAR takes into account wide-field effects (caused by the curvature of the sky), which are necessary given the $>30^\circ$ field of view of the MWA.

OSKAR is capable of beam-forming groups of receiver elements, such as those in an MWA tile. To be explicit, for these simulations, OSKAR was given the $4\times4$ grid pattern of an MWA tile, and told there was a cross-dipole antenna at each point. OSKAR then uses a analytic model for each cross-dipole and combines the response of all 16 antennas with appropriate delays to mimic the MWA primary beam. This is slightly different from the model use by the RTS~\citep{Sutinjo2014}, which includes mutual coupling between the dipoles.

OSKAR simulations were set up to exactly mimic the MWA observations detailed in~\ref{subsec:MWAdata}; the one difference being the correlator was set to sample at 2 seconds rather than the half second in the MWA data, which reduced the computational load by 4. The PUMA source list was used as an input sky model. When supplying the RTS with a calibration or peeling source list, the RTS reads in given flux densities at specified frequencies, and then extrapolates a sky model to the frequency of the data by fitting a power law spectral model between the two closest given frequencies. To ensure perfect source subtraction when using the PUMA source list positions, the sky model input to OSKAR was extrapolated in the same way to all frequencies.

A number of steps were necessary to run RTS on OSKAR simulations. OSKAR outputs either a native binary format file or a \texttt{CASA} measurement set\footnote{see \url{http://casa.nrao.edu/Memos/229.html}} (\texttt{MS}). The RTS is capable of reading either native MWA outputs or \texttt{UVFITS}\footnote{see Memo 117 \url{http://www.aips.nrao.edu/aipsmemo.html}} files. Routines already exist in \texttt{casapy} to transform a \texttt{MS} into a \texttt{UVFITS} file. Due to differing coordinate definitions and a frequency related issue however these \texttt{UVFITS} files still need editing (achieved here using \texttt{python}), leaving the final pipeline as:

\tikzstyle{box} = [rectangle, rounded corners, minimum height=1cm, minimum width=1cm, text centered, text width=1.3cm, draw=black, ]
\tikzstyle{vecArrow} = [thick, decoration={markings,mark=at position
   1 with {\arrow[semithick]{open triangle 60}}},
   double distance=1.4pt, shorten >= 5.5pt,
   preaction = {decorate},
   postaction = {draw,line width=1.4pt, white,shorten >= 4.5pt}]
   
\begin{figure}[h!]
\center
\begin{tikzpicture}[node distance=3cm]
\node (OSKms) [box] {OSKAR \texttt{MS}};
\node (OSKuv) [box, right of=OSKms] {OSKAR \texttt{UVFITS}};
\node (RTSuv) [box, right of=OSKuv] {RTS \texttt{UVFITS}};
 
\draw [vecArrow] (OSKms.350) -- node[anchor=east, xshift=0.65cm, yshift=0.25cm] {\texttt{casapy}} (OSKuv.190);
\draw [vecArrow] (OSKuv.350) -- node[anchor=east, xshift=0.65cm, yshift=0.3cm] {\texttt{python}} (RTSuv.190);

\end{tikzpicture}
\end{figure}

\begin{figure*}
\centering
\includegraphics[width=2.1\columnwidth]{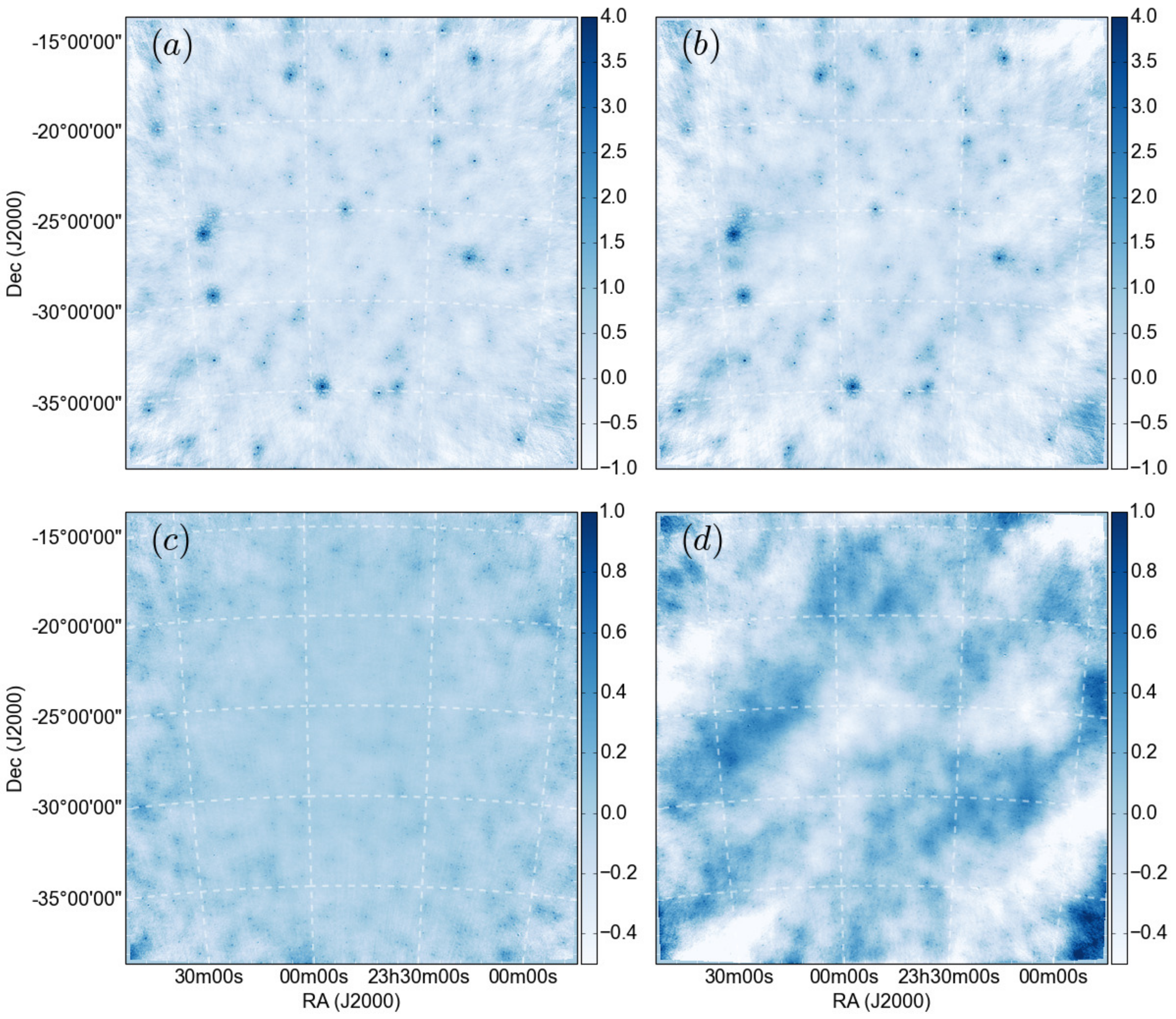}
\caption{\textsf{Four `dirty' (with the synthesized interferometric beam still convolved with the sky brightness) naturally weighted images are shown for an integration of 64s of data across the entire $30\,$MHz bandwidth. The left hand column shows OSKAR simulations, with the right hand showing MWA data. The top row shows calibrated data, and the bottom data with the same 1000 sources from the PUMA sky model created in \S\ref{sec:PUMAcat} subtracted. (a) and (b) reveal the excellent agreement of the synthesized beam created by OSKAR and the real MWA. (c) and (d) reveal the biggest difference in the sky, that being the diffuse emission clearly visible in (d); diffuse emission is mostly due to synchrotron emission from cosmic rays interacting local with galactic field lines~\citep{Ginzburg1969}.}}
\label{fig:OSKvsMWA}
\end{figure*}

One final complication is that the MWA correlator is slightly unusual in the fact that is does not fringe track - this being the procedure of adding phase delays within the signal path to ensure the data is fully coherent in the pointing direction~\citep[see Ch.2, ][for the theory of phase tracking]{Taylor1999}. The correlator in OSKAR is hard-coded to phase track, and so these phase rotations must be undone in the final \texttt{python} script. Furthermore, each simulated MWA tile is beam-formed in the direction of the phase centre, which is specified in $RA,\delta$. As the real MWA beam only points to a specific $HA,\delta$ during an observation, a new $RA,\delta$ must be entered for each time step. This combined with the frequency editing means OSKAR has to be run separately for every fine channel and time step. Given an input catalogue of 22,618 sources, it takes $\sim$6 hours using 24 NVIDIA Tesla C2090s GPUs\footnote{On the gSTAR super cluster \url{http://supercomputing.swin.edu.au/about-green-ii/}} to simulate a 2 minute MWA observation. A comparison between real MWA data and an OSKAR simulation is shown in Figure~\ref{fig:OSKvsMWA}.

\subsection{Results}
\label{subsec:results}

\begin{figure*}
\centering
\includegraphics[width=2.1\columnwidth]{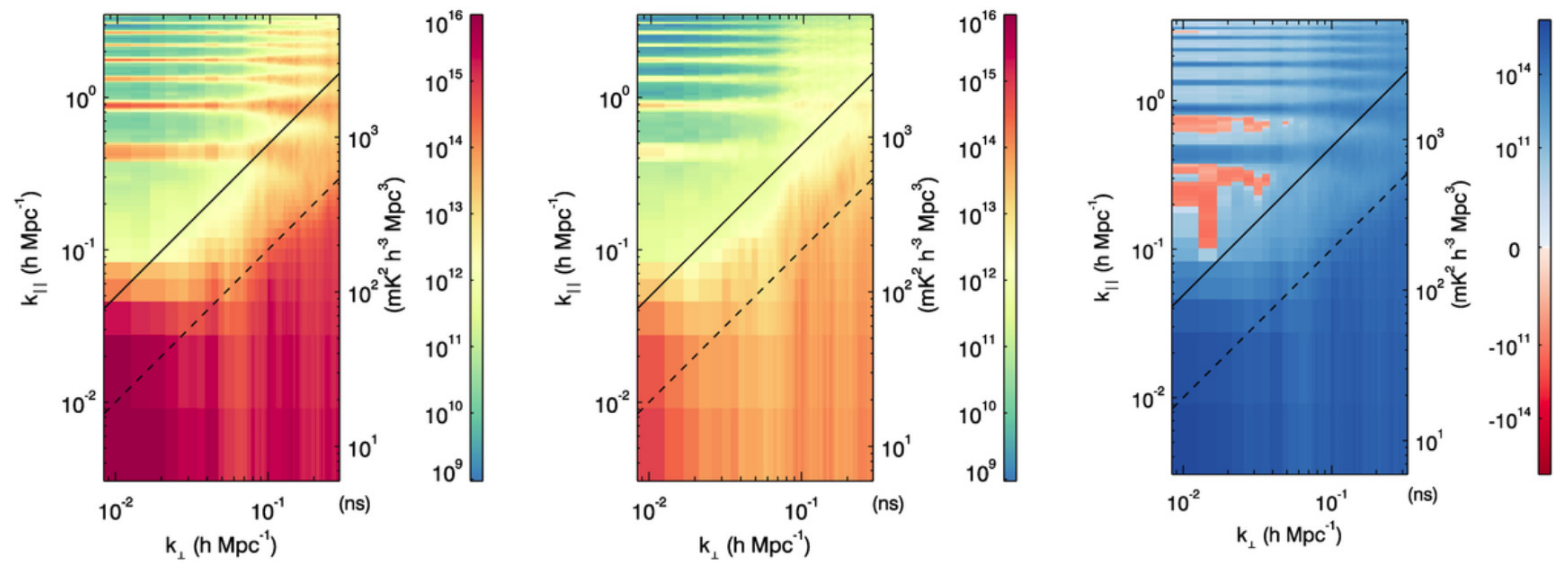}
\caption{\textsf{Two 2D power spectra are shown (left and centre), both created using the XX polarisation and the entire simulated hour of data. Each plot shows amplitude as a function of $k$-modes perpendicular to the line of sight (derived from angular scales on the sky, $k_{\perp}$) horizontally, and $k$-modes parallel to the line of sight (derived from frequency response, $k_{\parallel}$) vertically. The plot on the left shows the power before source subtraction, and the centre after 1000 sources have been subtracted. The plot on the right shows the difference plot of the 2D power spectra, with the 1000 source spectra subtracted from the spectra without source subtraction. Blue in this case shows more power being present before source subtraction. The absolute scale shown here is not the most instructive part of these plots as an interferometer naturally measures variations about a mean; the relative power as a function of $k$-space however informs us where foreground power is being removed from.}}
\label{fig:CHIPS}
\end{figure*}

The left and central plots in Figure~\ref{fig:CHIPS} show 2D PS obtained after processing the OSKAR simulations through the Australian MWA pipeline. Each plot shows the power as a function of $k$-modes derived from angular scales upon the sky ($k_{\perp}$) and those derived from the frequency response of the instrument ($k_{\parallel}$). The solid diagonal line is set to represent the expected location of foregrounds within the main lobe of the MWA primary beam, and the dashed diagonal line the contribution of foreground sources at the horizon limit.

Given that the visibilities only contain point sources the calibration solutions were excellent when calibrating both with the PUMA and MWACS source lists. The PS in Figure~\ref{fig:CHIPS} display the characteristic wedge shaped power from point sources, apparent in the saturated bottom right corner of each plot. The horizontal bands of power in the EoR window are due to the spectral behaviour of the flagged fine channels (see \S\ref{subsec:MWAdata}) convolving with the power from the wedge. When searching for calibration or data analysis effects in the 2D PS, it is often instructive to create difference plots, that being one 2D PS subtracted from another~\citep[e.g.][in press]{Beardsley2016}. An example difference plot is shown on the right in Figure~\ref{fig:CHIPS}.

After processing through CHIPS with various subsets of the simulated data to demonstrate the effect of subtracting sources with positional inaccuracies, we discovered a competing differencing effect inherent to the data: each pointing observed a fundamentally different sky. For this reason, we split our results into zenith and off-zenith pointings. We include a comparison of zenith and off-zenith simulated data in this section for completeness.

\begin{figure*}
\centering
	\begin{subfigure}{0.4\textwidth}
	\includegraphics[width=\textwidth]{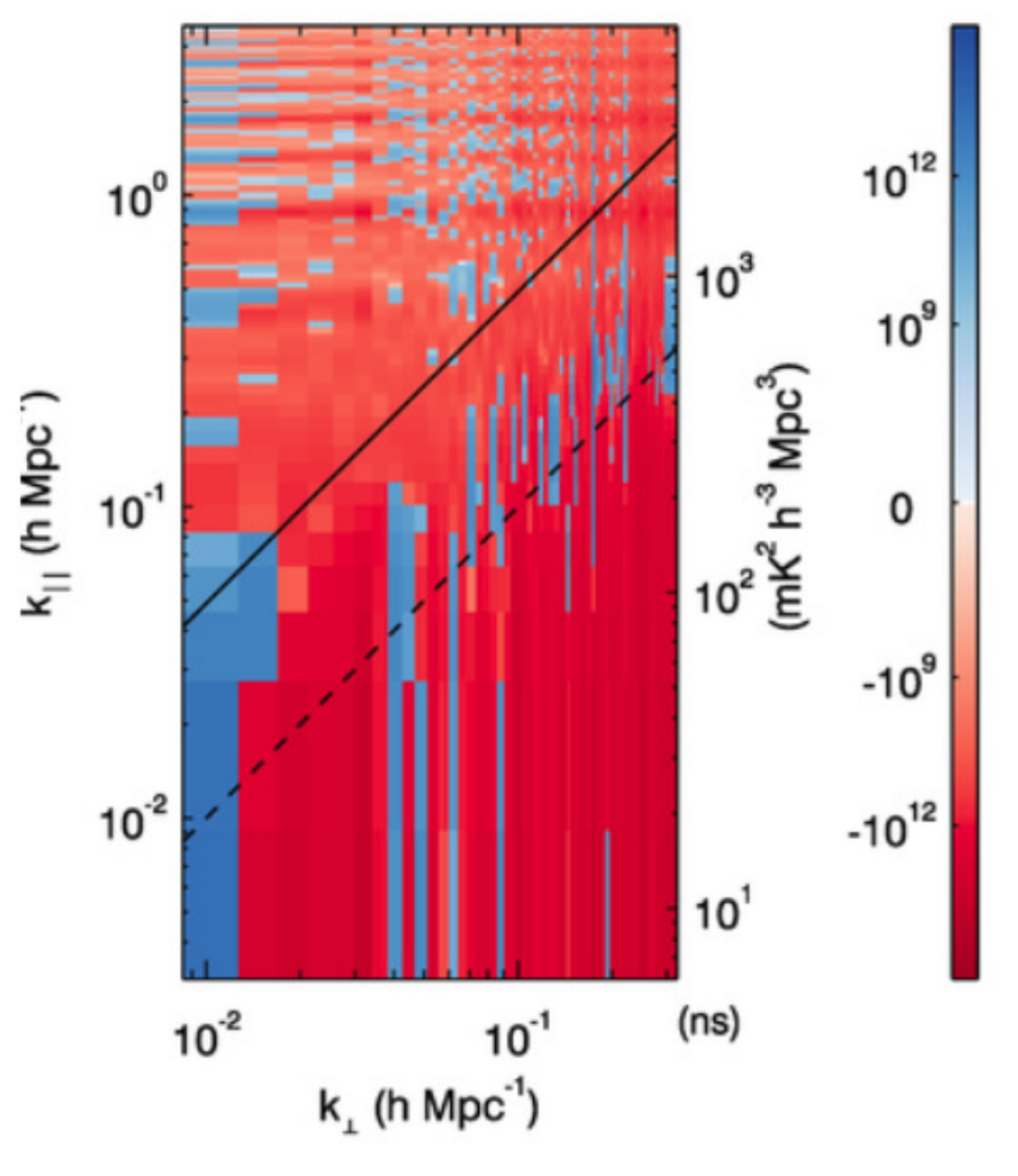}
	\caption{\textsf{Zenith pointing, PUMA source list $-$ MWACS source list, XX}}
	\label{subfig:complistA}
	\end{subfigure}
	\qquad
	\begin{subfigure}{0.4\textwidth}
	\includegraphics[width=\textwidth]{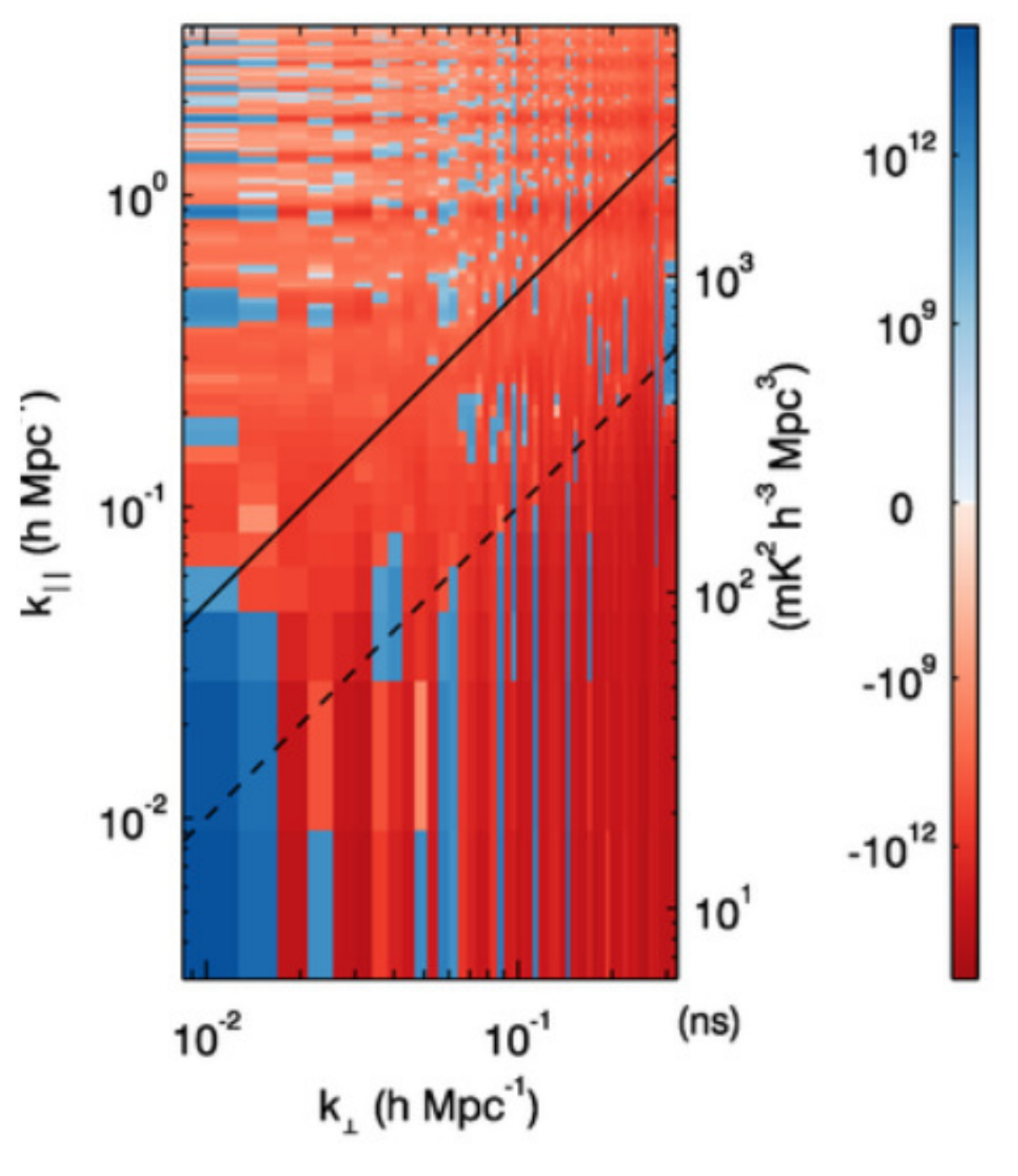}
	\caption{\textsf{Zenith pointing, PUMA source list $-$ MWACS source list, YY}}
	\label{subfig:complistB}
	\end{subfigure}
	
	\begin{subfigure}{0.4\textwidth}
	\includegraphics[width=\textwidth]{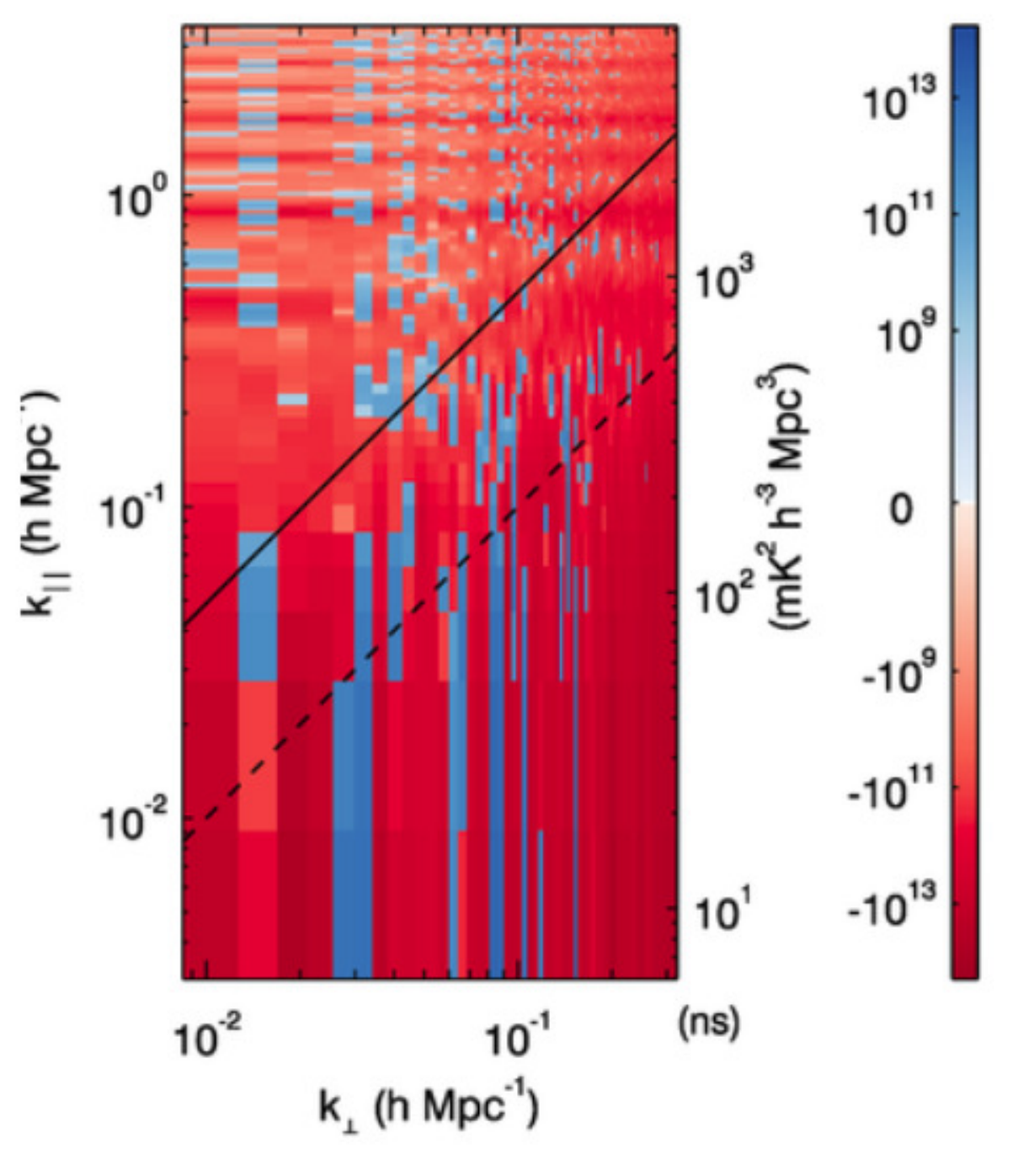}
	\caption{\textsf{Off-zenith pointing, PUMA source list $-$ MWACS source list, XX}}
	\label{subfig:complistC}
	\end{subfigure}
	\qquad
	\begin{subfigure}{0.4\textwidth}
	\includegraphics[width=\textwidth]{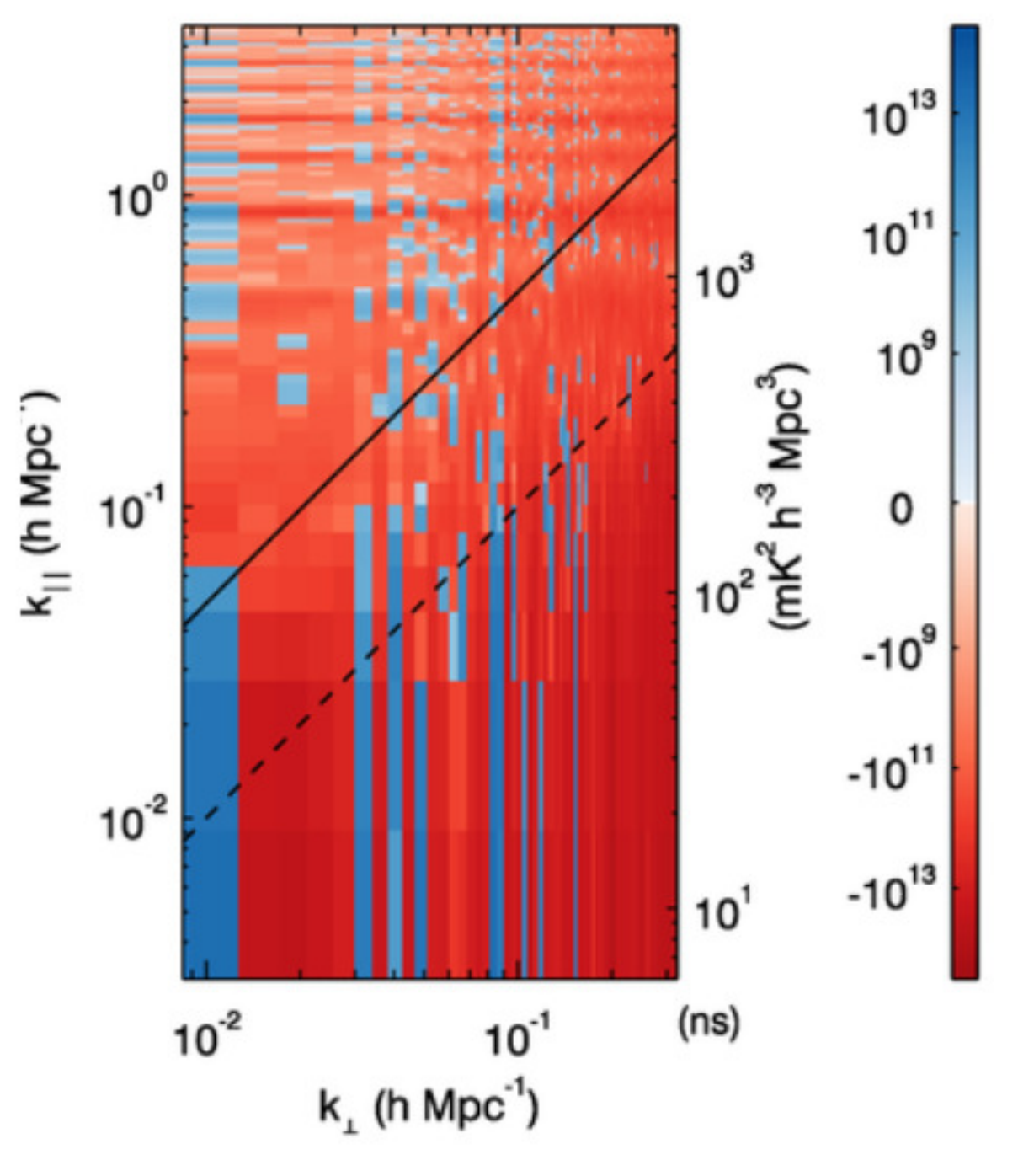}
	\caption{\textsf{Off-zenith pointing, PUMA source list $-$ MWACS source list, YY}}
	\label{subfig:complistD}
	\end{subfigure}
\caption{\textsf{4 difference PS are shown to contrast data processed with the PUMA source list to the MWACS source list. In each case, blue represents more power for data with exact positional source subtraction opposed to offset positional subtraction, and red less power.}}
\label{fig:compare_list}
\end{figure*}

The effect of subtracting sources with positional inaccuracies is shown in Figure~\ref{fig:compare_list}. This shows the difference between subtracting sources from the PUMA and MWACS source list. The PS are compared pointing by pointing to negate the pointing effect described above. The behaviours observed at low $k_{\perp}$, $k_{\parallel}$ are inconsistent, which is likely due to the poor sampling that exists here; this part of the PS corresponds to large angular and spectral scales, which are limited due to the field of view and bandpass. For the rest of the PS, it is seen consistently over each pointing and polarisation that subtraction with exact source positions not only overall removes more power from the wedge, but also from the window as well. As stated, the absolute value of the power in the 2D PS shown here are not instructive, but the relative differences in power seen in the window are within~$\sim$1-2 orders of magnitudes of that seen in Figure~\ref{fig:CHIPS}\footnote{It should be noted however that these PS are made with only half an hour of data, so these orders of magnitude should be taken as indicative only; an actual detection requires on order of~$\sim$1000 hours~\citep{Beardsley2013}}. Given that the EoR signal is thought to be~$\sim$3 orders of magnitude dimmer than these foregrounds, this highlights the need for the best possible foreground models to reduce the leakage of point sources from the wedge into the window~\citep[i.e.][]{Offringa2016}.

\begin{figure*}
\centering
	\begin{subfigure}{0.4\textwidth}
	\includegraphics[width=\textwidth]{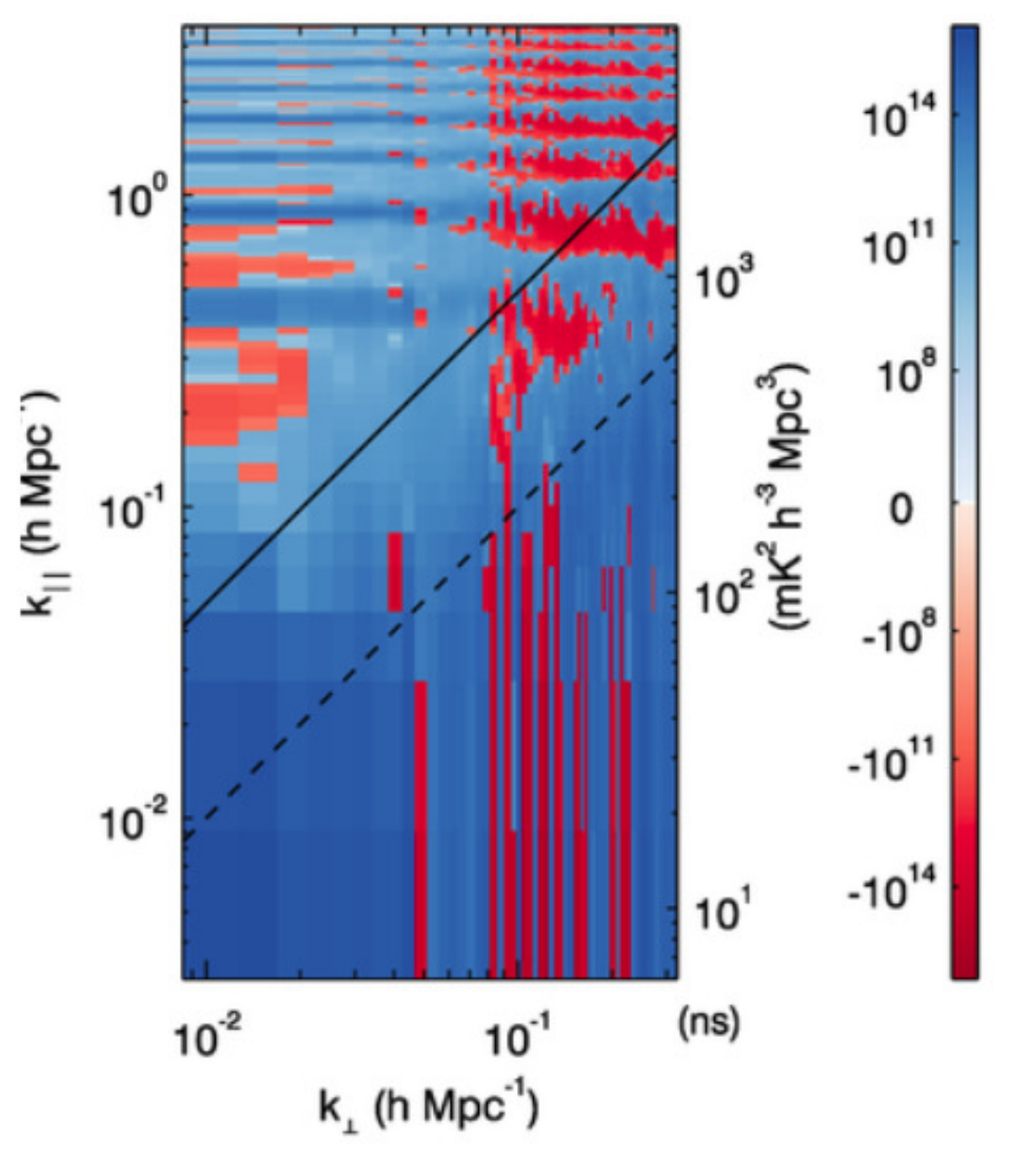}
	\caption{\textsf{No peeling, zenith $-$ off-zenith, XX polarisation}}
	\label{subfig:comppointA}
	\end{subfigure}
	\qquad
	\begin{subfigure}{0.4\textwidth}
	\includegraphics[width=\textwidth]{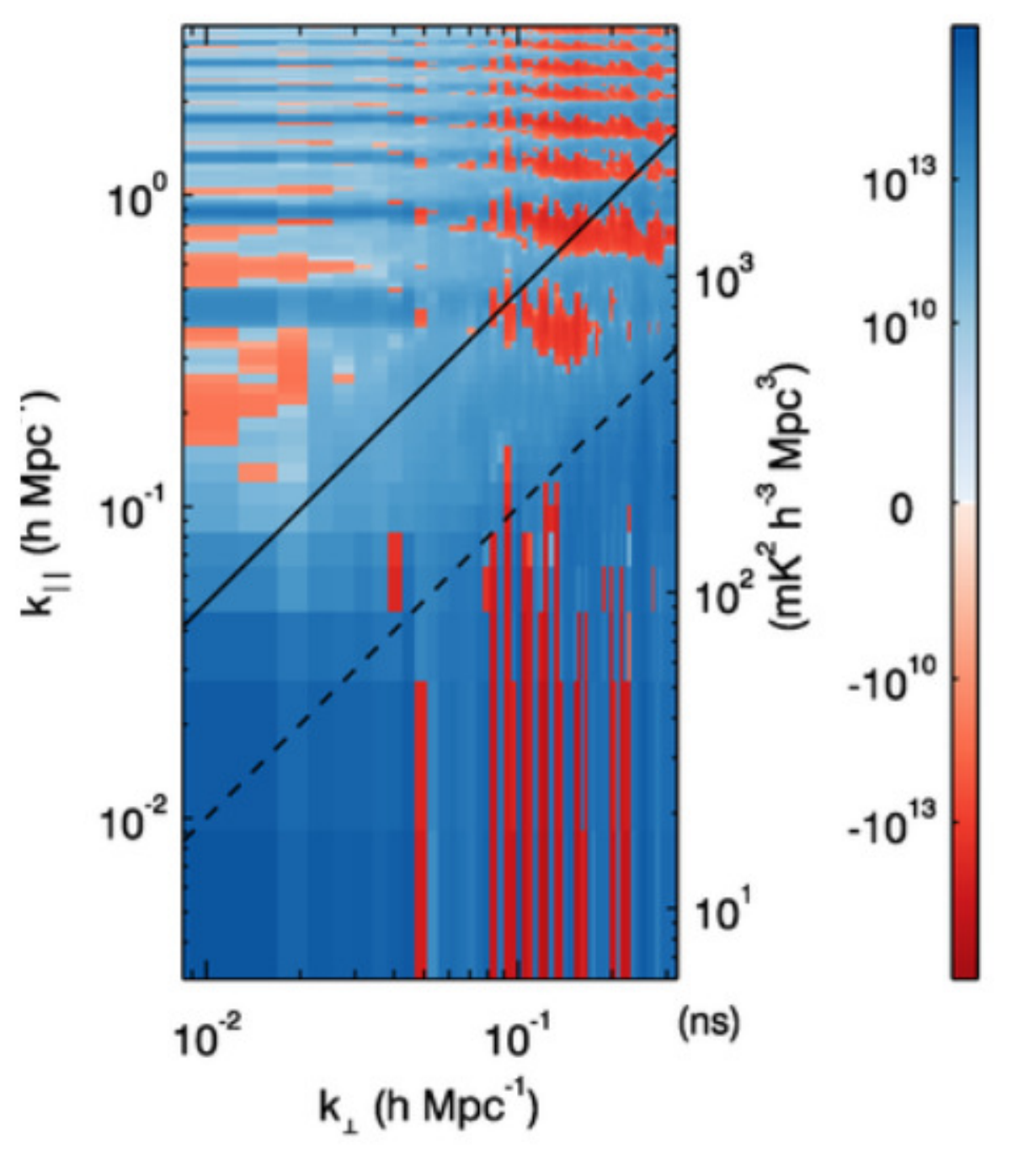}
	\caption{\textsf{No peeling, zenith $-$ off-zenith, YY polarisation}}
	\label{subfig:comppointB}
	\end{subfigure}
	
	\begin{subfigure}{0.4\textwidth}
	\includegraphics[width=\textwidth]{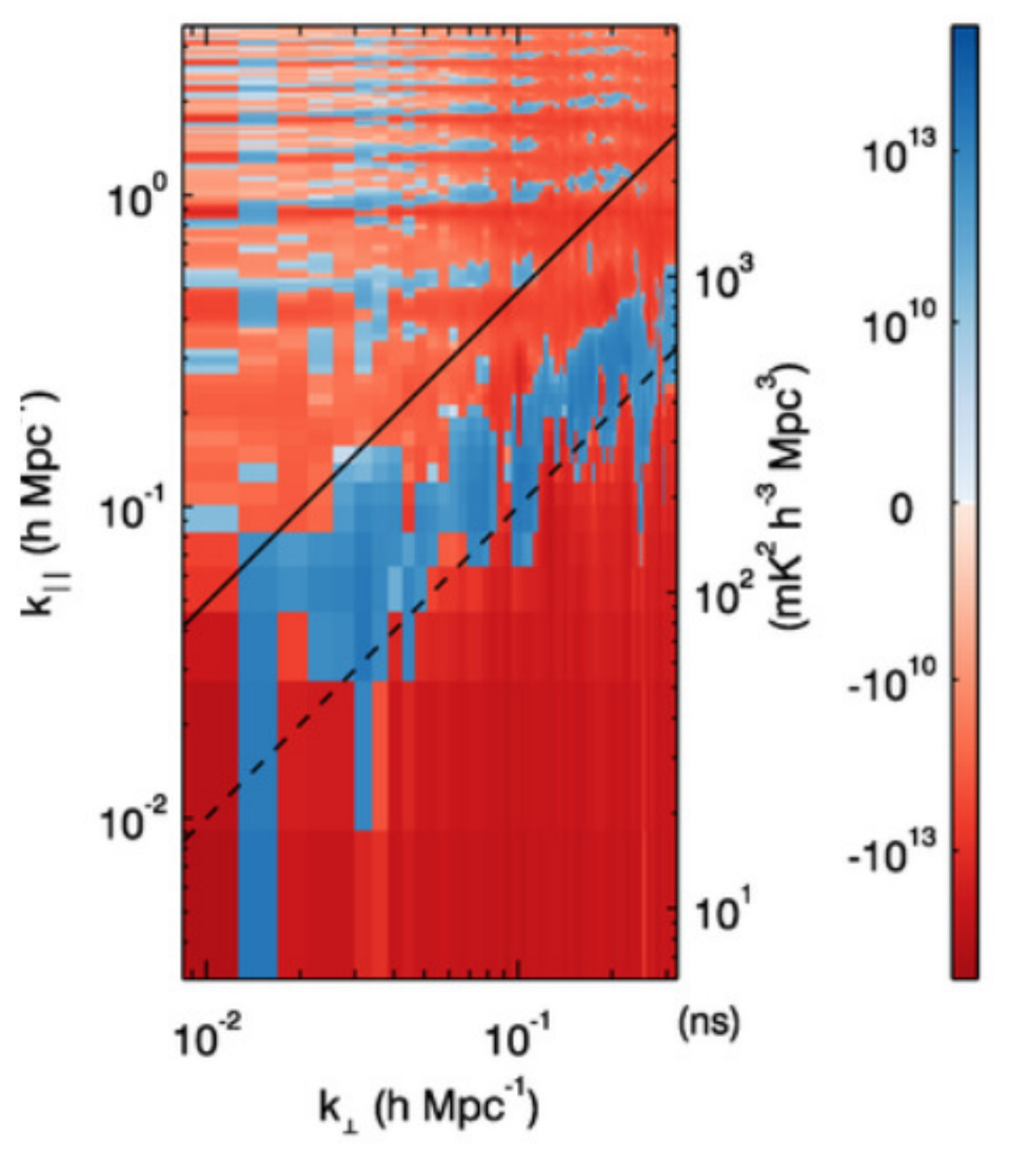}
	\caption{\textsf{1000 sources peeled, zenith $-$ off-zenith, XX polarisation}}
	\label{subfig:comppointC}
	\end{subfigure}
	\qquad
	\begin{subfigure}{0.4\textwidth}
	\includegraphics[width=\textwidth]{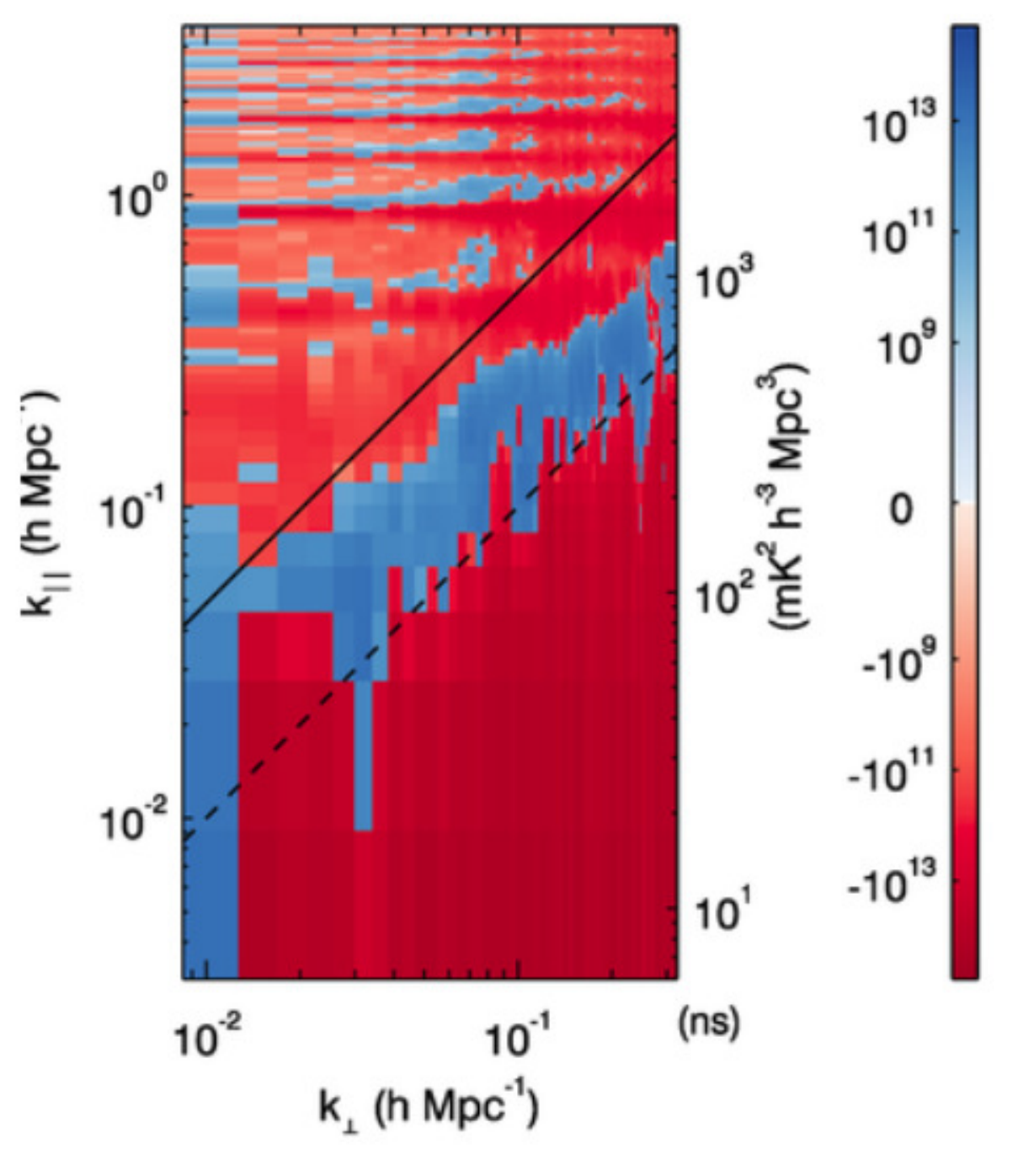}
	\caption{\textsf{1000 sources peeled, zenith $-$ off-zenith, YY polarisation}}
	\label{subfig:comppointD}
	\end{subfigure}
\caption{\textsf{4 difference PS are shown for data processed using the PUMA source list only, each representing half an hour of data. Each sub-caption details which data are shown; in each case, blue represents more power in a zenith pointing, and red less power. The top row shows there is overall more power seen for the zenith pointing before source subtraction, and the bottom rows shows there is overall less power after subtracting the 1000 brightest sources. Again, the absolute value of the power is less important than the distribution of power throughout $k$-space.}}
\label{fig:compare_pointing}
\end{figure*}

The differences in the sky seen by differing pointings is shown in Figure~\ref{fig:compare_pointing}. These plots demonstrate that the sky is fundamentally sampled differently through the zenith and off-zenith pointings, due to a combination of the beam pattern differing between pointings, creating grating side lobes which contribute power at different amplitudes from different parts of the sky, along with a sky that is changing with time. This is perhaps most apparent in Figures~\ref{subfig:comppointC} and~\ref{subfig:comppointD}, shown by the diagonal blue stripe bound by the solid and dashed lines. Power in this area of the PS comes from sources close to the horizon~\citep[see][and references therein]{Pober2016}; the differences caused here by the changed horizon between pointings become apparent after subtracting the 1000 brightest sources.

\section{Discussion}
\label{sec:discuss}

We have developed software that is capable of automatically cross-matching~$\sim$99\% of the MWACS catalogue to other radio-wavelength catalogues of differing resolutions, necessitating the need to deal with confused matching. Using simplistic simulations, we tested the matching results of PUMA and found it able to reliably recover a known SI distribution, and to be robust to typical ionospheric positional offsets prevalent in low radio-frequency observations. This high fidelity of matching is important for the current generation of low radio-frequency surveys (such as GLEAM and MSSS), which will produce catalogues on orders of~$\sim10^5$ sources, and more so for future SKA-type surveys which may produce~$\sim10^6$. With this in mind, the software has been designed to be tunable to the needs of the desired application, with several manually adjustable parameters. Even so, these algorithms will need to be further developed to save manually inspecting over $10^4$  sources\footnote{Alternatively, these difficult cross-matches could be opened up as a citizen science project, akin to the Radio Galaxy Zoo -  \url{https://www.zooniverse.org/}. The details of setting this up would not be straight forward however.}(although these numbers of sources are routinely inspected spread over a large team for a wide-field optical survey).

After applying PUMA, the impact of gaining more precise positional information from higher-frequency radio catalogues was investigated through OSKAR simulations. This was achieved by studying the effects of removing sources from interferometric visibilities and investigating the resulting 2D PS. It was found that when subtracting sources with exactly correct positions, more power was not only subtracted from the wedge but also from the window, the measurement space in which an EoR detection could potentially be made. This adds weight to the growing argument in the literature to the most accurate possible sky models.

While this paper concentrates on the benefits of this methodology for creating foreground models for EoR science, it of course has wider applications, particularly for population studies of radio galaxies, and for verification of sources during catalogue creation. It also has implications for baseline configurations for future EoR arrays: if the true positions of sources can be established from higher frequency information, longer baselines may not be necessary, reducing cost and allowing for more short baselines, increasing the sensitivity of the array to the spatial scales at which the EoR signal can be measured.

\section*{Acknowledgements}

J.~L~.B.~L. wishes to thank the anonymous referee for inspiring \S\ref{sec:testPUMA}, and providing many other valuable suggestions that greatly improved the paper. J.~L~.B.~L. wishes to acknowledge the support of the MIR and MIFR scholarships afforded by the University of Melbourne. This work was supported by resources awarded under Astronomy Australia Ltd’s merit allocation scheme on the gSTAR national facility at Swinburne University of Technology. gSTAR is funded by Swinburne and the Australian Government’s Education Investment Fund. This work was also supported by resources provided by the Pawsey Supercomputing Centre with funding from the Australian Government and the Government of Western Australia. The simulations in this work make use of the Murchison Radio-astronomy Observatory, operated by CSIRO. We acknowledge the Wajarri Yamatji people as the traditional owners of the Observatory site. Support for the operation of the MWA is provided by the Australian Government Department of Industry and Science and Department of Education (National Collaborative Research Infrastructure Strategy: NCRIS), under a contract to Curtin University administered by Astronomy Australia Limited. We acknowledge the iVEC Petabyte Data Store and the Initiative in Innovative Computing and the CUDA Center for Excellence sponsored by NVIDIA at Harvard University. This research was conducted by the Australian Research Council Centre of Excellence for All-sky Astrophysics (CAASTRO), through project number CE110001020.


\bibliographystyle{pasa}
\bibliography{edited_library}


\appendix
\section{Algorithms}
\label{app_Alg}

\begin{algorithm}
  Record the number of combinations \;
  Work out which catalogues are repeated \;
  Create retained\_{}combinations list \;
  Calculate the distance of the repeated sources to the base catalogue \;
  \For{each combination i}
    {
    \uIf{$P(H|D_i) > P_{u}$}
      {accept combination; append to retained\_{}combinations} 
    \uElseIf{source from repeated catalogue is within resolution of base catalogue source + error}
        {accept combination; append to retained\_{}combinations}
    \Else{reject combination}  
    }
  Record the number of retained combinations \;
  \eIf{number of retained combinations == 0}{
       reject group; label as rejected positionally
       }{
       return retained\_{}combinations
       }
       
  \caption{\sffamily{Positional selection criteria for all cross-match combinations associated with a base source. Any catalogue with more than one match source is labelled as `repeated'. The algorithm accepts a combination if it is either likely, or if the repeated source is within the resolution of the base catalogue. The retained combinations are then investigated through Algorithms~\ref{spec_dom_crit} and ~\ref{comb_crit_test}. $P_u$ can be modified by the user. At all stages, statistics of the matching process are gathered to propagate through to the final matched catalogue.}}
\label{pos_crit_multi}
\end{algorithm}
\begin{algorithm*}
  Fit a power law to the spectral data \;
  Calculate the residuals of fit \;
  \eIf{$P(H|D) > P_{u}$}
    {
    label as \texttt{isolated} match \;
    accept combination
    }{
       Calculate the distance of all matched sources to the base catalogue \;
       \eIf{all sources within resolution of base catalogue source + error}
           {
           \eIf{$\epsilon^2 > \epsilon^2_u$ and $\chi_{red}^2 > \chi_{red,u}^2$}
             {
             reject combination\; 
             label as rejected spectrally
             }{
             label as \texttt{isolated}\;
             }
           }{
           reject combination; label as rejected positionally
           }
     }
  \caption{\sffamily{Positional selection criteria for a single source cross-match. If there is only one combination possible, and it has a positional probability over a given threshold, it is accepted without scrutinizing the spectral data. This avoids assuming any spectral model. If the match is below $P_u$ , all matched sources are checked to be within the resolution of the base catalogue. As there was only one possible match, a high positional probability was expected, so a spectral test is applied. If the residuals $\epsilon$, $\chi_{red}^2$ of a fit to a power law (as detailed in \S\ref{subsec_IsoMatch}) are below a certain threshold $\epsilon^2_u$, $\chi_{red,u}^2$, the source is accepted. At all stages, statistics of the matching process are gathered to propagate through to the final matched catalogue.}}
\label{pos_crit_single}
\end{algorithm*}

\begin{algorithm*}
  create string pos\_{}dom = ``none" \;
  create string spec\_{}dom = ``none" \;
  \For{each combination i}{
    calculate residual ratio $\epsilon_i$ / $\epsilon_j$ where $i\neq j$ \;
    calculate residual ratio $\chi_{red,i}^2$ / $\chi_{red,j}^2$ where $i\neq j$ \;
    
    \eIf{all residual ratios are $<=$ 0.33 for either $\epsilon$ or $\chi_{red}^2$} 
      {
      spec\_{}dom = ``combination i" \tcc*[f]{This means the residuals for this combination are at least 3 times smaller than all other combinations}
      }{
      pass
      }
   
    \eIf{$P(H|D_i) > P_u$ and all $P(H|D_j) < P_l$ where $i\neq j$}{
      pos\_{}dom = ``combination i" \tcc*[f]{This means that this one combination is likely and all others are unlikely}
      }{
      pass
     }
   }
  \eIf(\tcc*[f]{This means combination i is positionally most likely and spectrally dominant}){pos\_{}dom == spec\_{}dom}{
    label combination i as \texttt{dominant} \;
    accept combination i
    }{
    pass retained\_{}combinations on to Algorithm~\ref{comb_crit_test}
  }
  
  \caption{\sffamily{A test for spectral dominance. If one combination has residuals that are at least 3 times smaller than all other combinations, and is positionally likely whilst all other combinations are unlikely, accept the source. Positional and spectral dominance are required at the same time, to rule out chance alignment of sources with particular flux densities. Otherwise, the combinations are passed on to Algorithm~\ref{comb_crit_test}. At all stages, statistics of the matching process are gathered to propagate through to the final matched catalogue.}}
\label{spec_dom_crit}
\end{algorithm*}

\begin{algorithm*}
  combine the flux densities of the source from the same catalogue \;
  combine the flux density errors \;
  calculate $\epsilon^2$ with the new combined flux density and error\;

  \eIf{ $\epsilon^2 <= \epsilon^2_u$ or $\chi_{red}^2 <= \chi_{red,u}^2$ }
    {
	\eIf{distance between repeated sources $> d_{split}$, and splitting implemented}
	  {
	  send to Algorithm~\ref{split_crit_test}
	  }{
      create a weighted RA, Dec and errors as described in Eq.~\ref{new_RA} \;
      label as \texttt{multiple} \;
      accept combination with updated position, flux density and errors \;
      }
    }{
    send source information to be investigated by eye; label as retained after combining
    }
   
   \caption{\sffamily{A test for source combining. If no one combination passes Algorithm~\ref{spec_dom_crit}, try combing the flux densities from the sources from the same catalogue. If the combined flux densities pass a spectral test, create a new position for the combined source, weighting the RA and Dec of each source by its flux density. If splitting is implemented, pass to Algortihm~\ref{split_crit_test}. Otherwise accept the combined source. If the combination of flux densities does not pass, send the combinations to be investigated by eye. At all stages, statistics of the matching process are gathered to propagate through to the final matched catalogue.}}
\label{comb_crit_test}
\end{algorithm*}

\begin{algorithm*}
  count number of repeated sources for each catalogue \;
  \eIf{all repeated catalogues have the same number of sources}
    {
    match the components of the repeated catalogues \;
    split the flux density of the single catalogues as described in Eq.~\ref{avg_weight} \;
    spectrally test each new component as described in Algorithm~\ref{spec_dom_crit} \;
	\eIf{all components pass spectral test}
	   {
	   accept all components
	   }{
	   send source information to be investigated by eye; label as retained after splitting
	   }    
    }{
    return info to Algorithm~\ref{comb_crit_test} for combinational test
    }
   
   \caption{\sffamily{A test for source splitting. If a source can be combined, but the components to be combined are separated by a distance larger than the user specified $d_{split}$, the combination is tested for splitting. If more than one catalogue has repeated sources, the Algorithm requires they have the same amount of sources. Each set of repeated sources are then matched by distance to create components. An SED is constructed for each component, and fit to the linear model. If all components pass the spectral test, the cross-match combination is split up in to multiple cross-matched sources.}}
\label{split_crit_test}
\end{algorithm*}

\appendix
\section{Parameter Space}
\label{app:params}
\begin{figure*}
\centering
	\begin{subfigure}{2.0\columnwidth}
	\includegraphics[width=\textwidth]{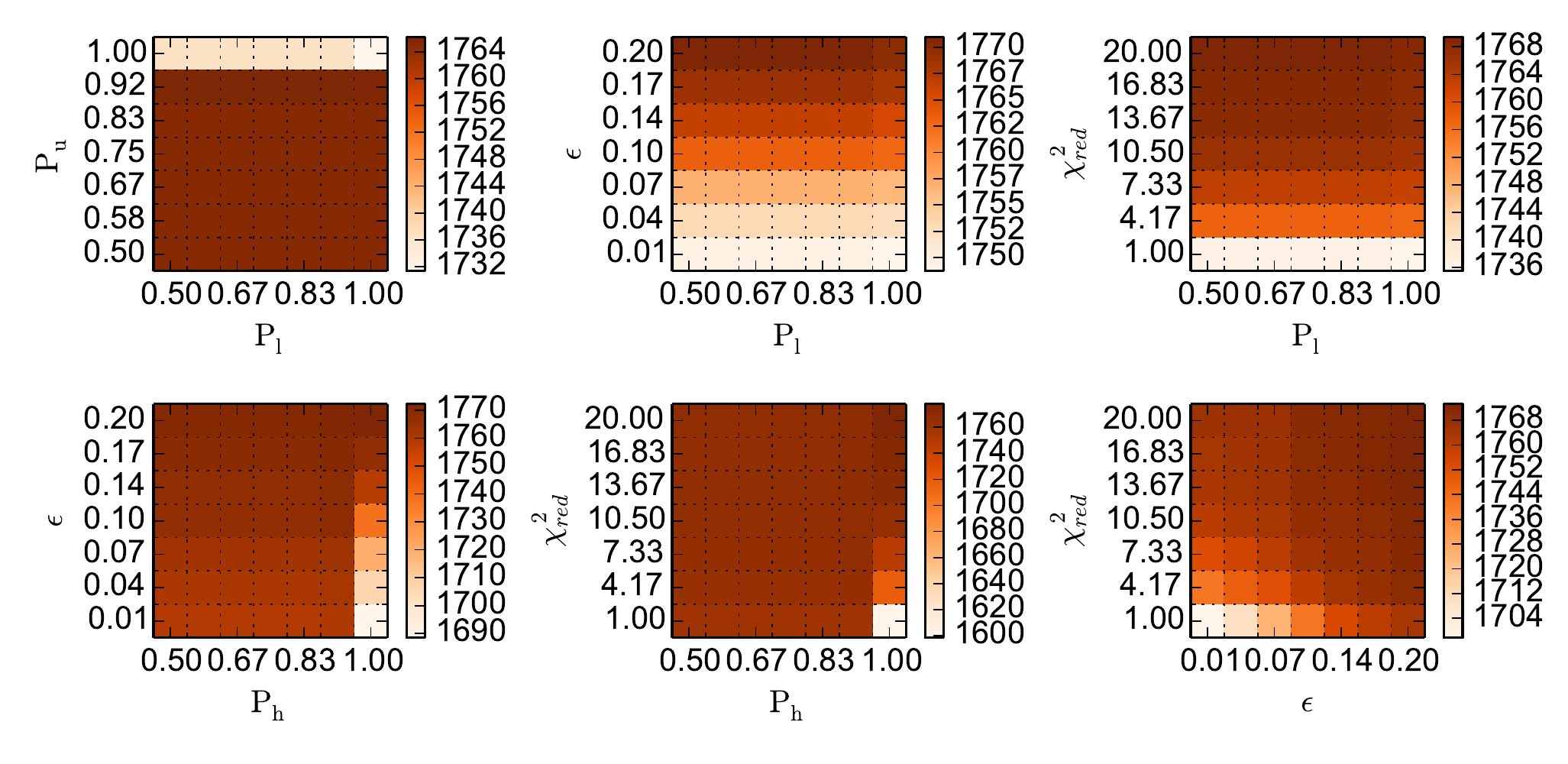}
	\caption{\textsf{\texttt{isolated} cases}}
	\label{subfig:isolated}
	\end{subfigure}
	
	\begin{subfigure}{2.0\columnwidth}
	\includegraphics[width=\textwidth]{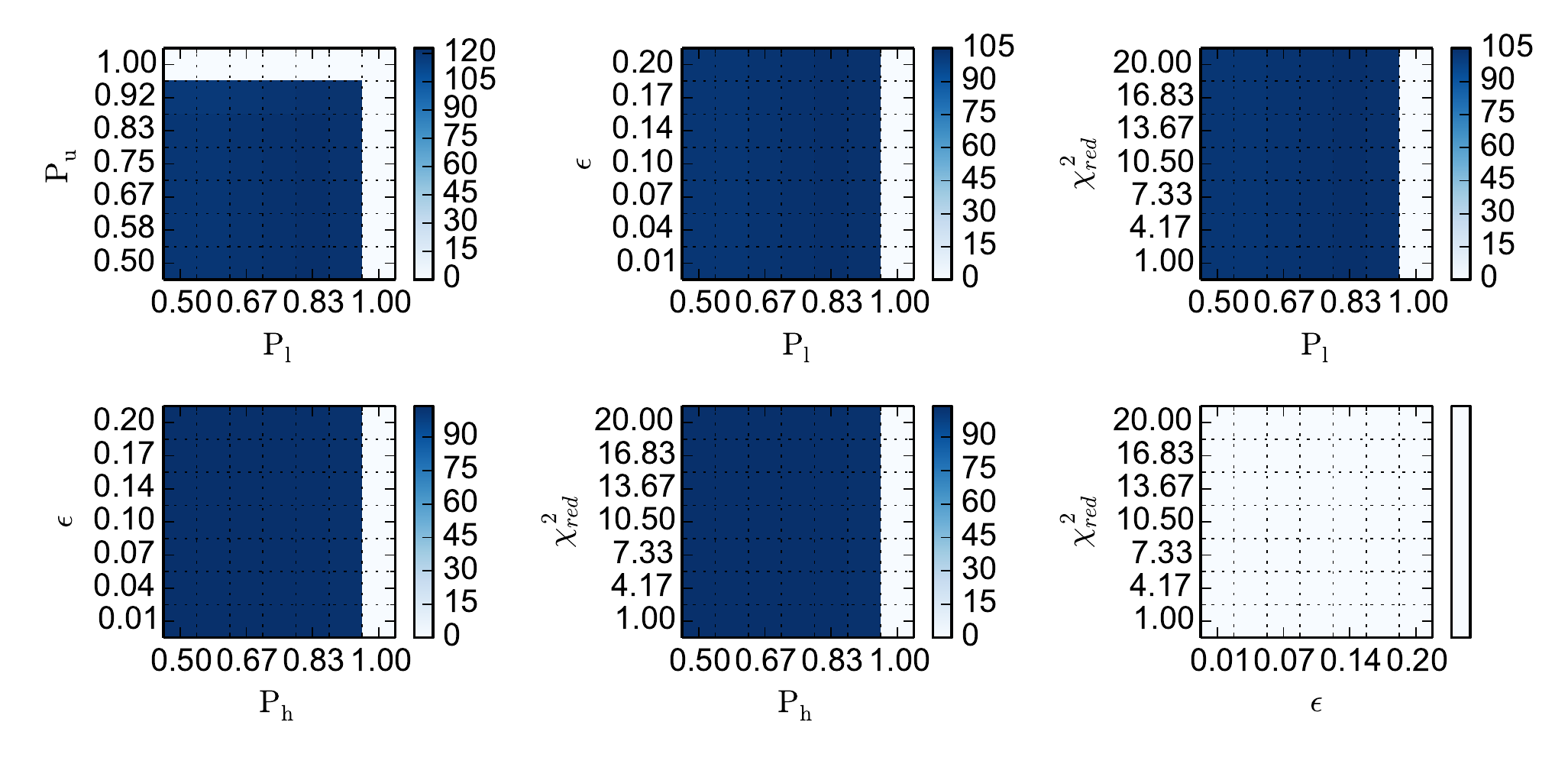}
	\caption{\textsf{\texttt{dominant} cases}}
	\label{subfig:dominant}
	\end{subfigure}
\caption{\textsf{An exploration of the effects of parameter space for \texttt{isolated} and \texttt{dominant} classifications. The bottom right panel of \ref{subfig:isolated} shows that $\chi^2_{red}$ and $\epsilon$ have no effect on the number of dominant sources; this is because dominance is established using a ratio of residuals, rather than a cut-off.}}
\label{fig:params-iso-dom}
\end{figure*}

\begin{figure*}
\centering
	\begin{subfigure}{2.0\columnwidth}
	\includegraphics[width=\textwidth]{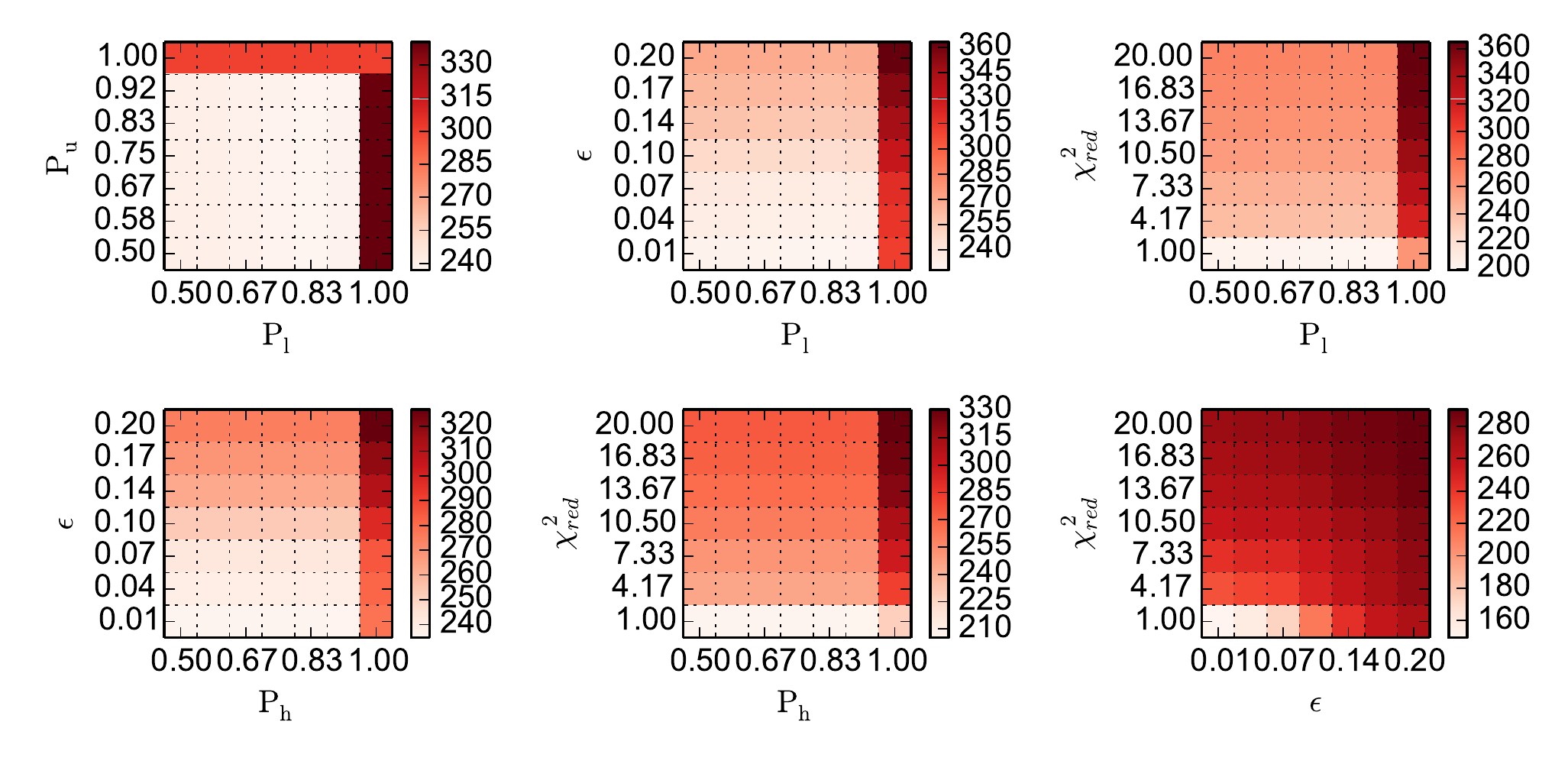}
	\caption{\textsf{\texttt{multiple} cases}}
	\label{subfig:multiple}
	\end{subfigure}
	
	\begin{subfigure}{2.0\columnwidth}
	\includegraphics[width=\textwidth]{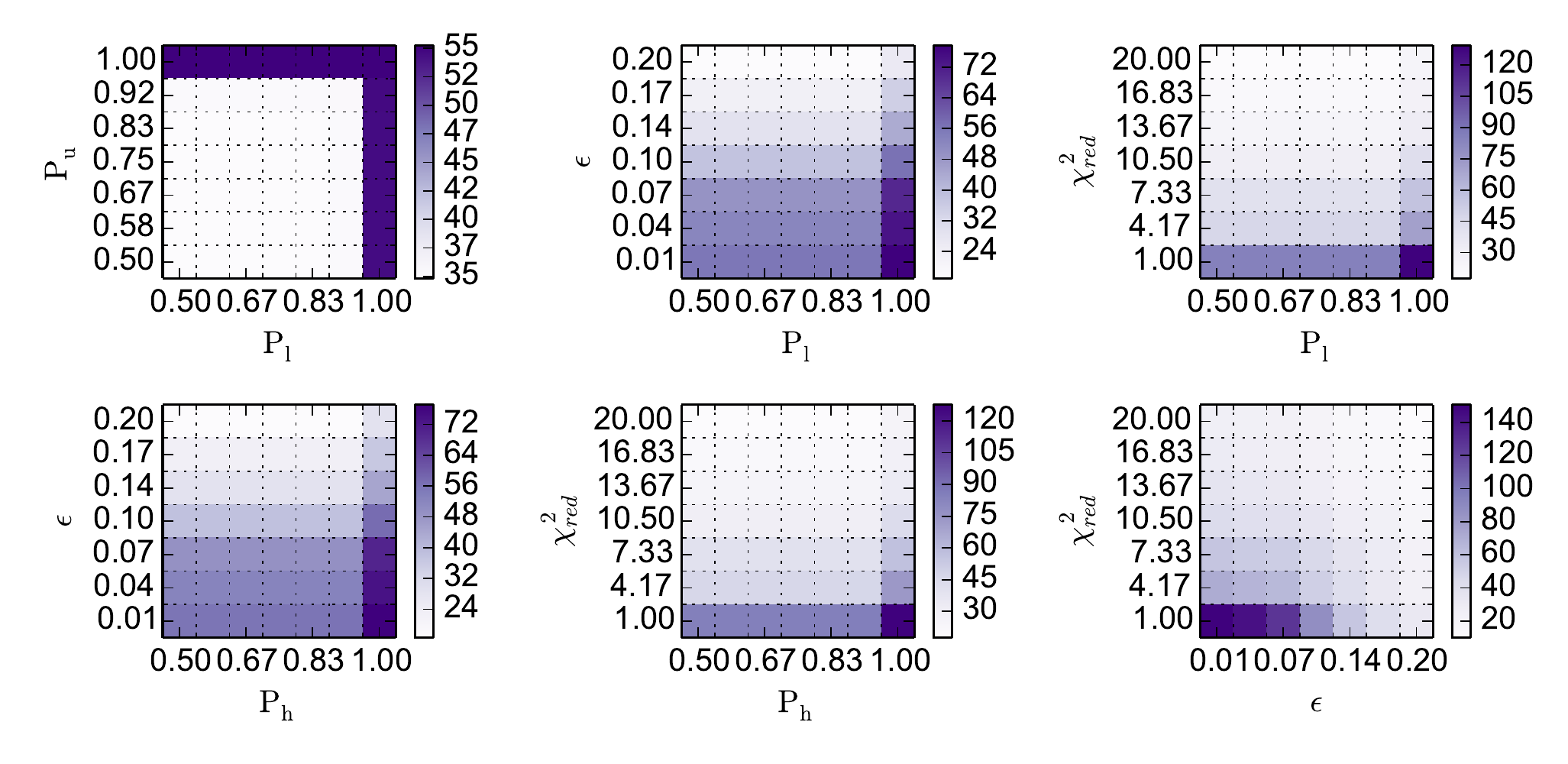}
	\caption{\textsf{\texttt{eyeball} cases}}
	\label{subfig:eyeball}
	\end{subfigure}
\caption{\textsf{An exploration of the effects of parameter space for \texttt{multiple} and \texttt{eyeball} classifications.}}
\label{fig:params-mul-eye}
\end{figure*}

\begin{figure*}
\centering
	\begin{subfigure}{2.0\columnwidth}
	\includegraphics[width=\textwidth]{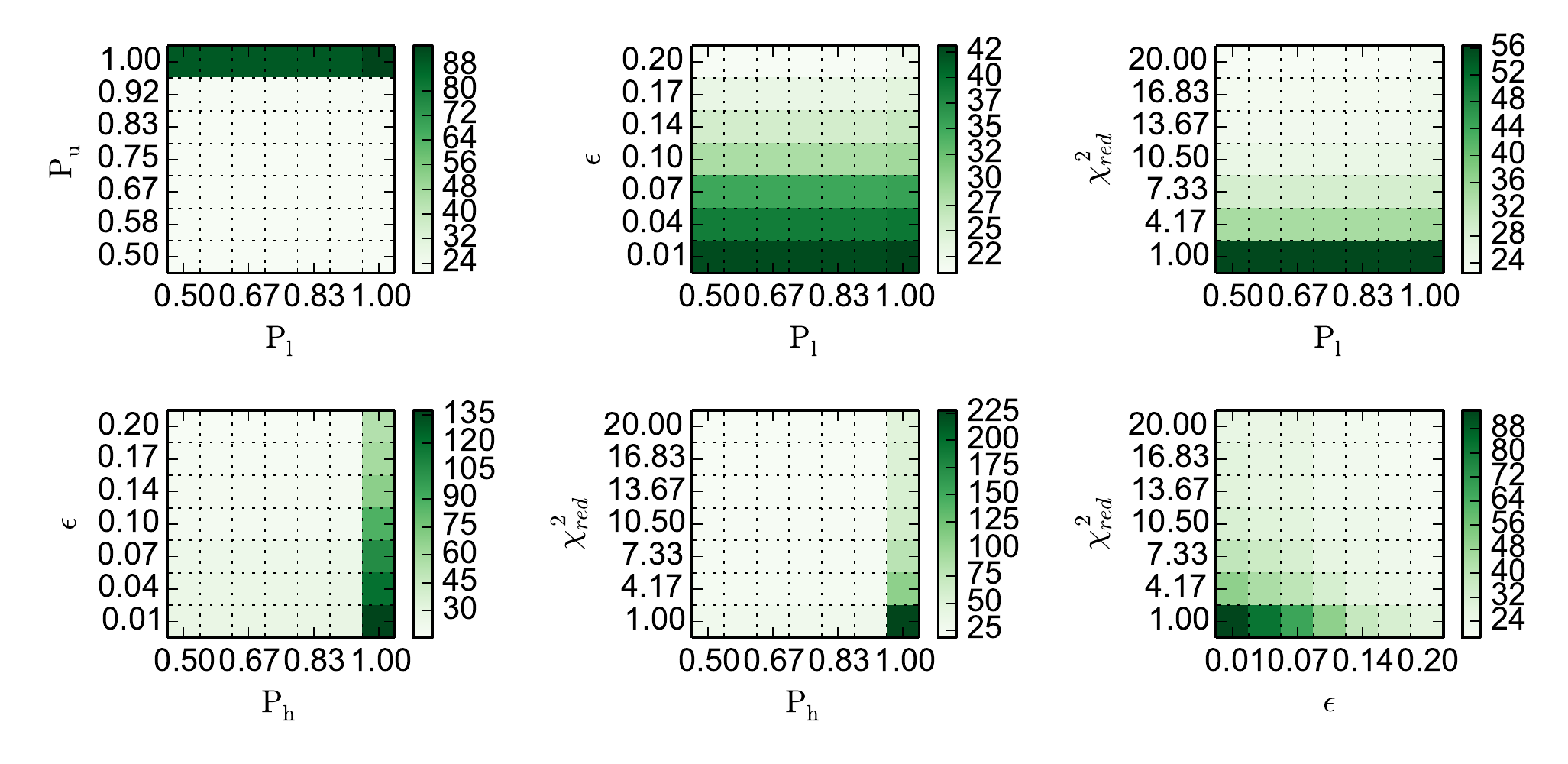}
	\caption{\textsf{\texttt{reject} cases}}
	\label{subfig:reject}
	\end{subfigure}
	
	\begin{subfigure}{2.0\columnwidth}
	\includegraphics[width=\textwidth]{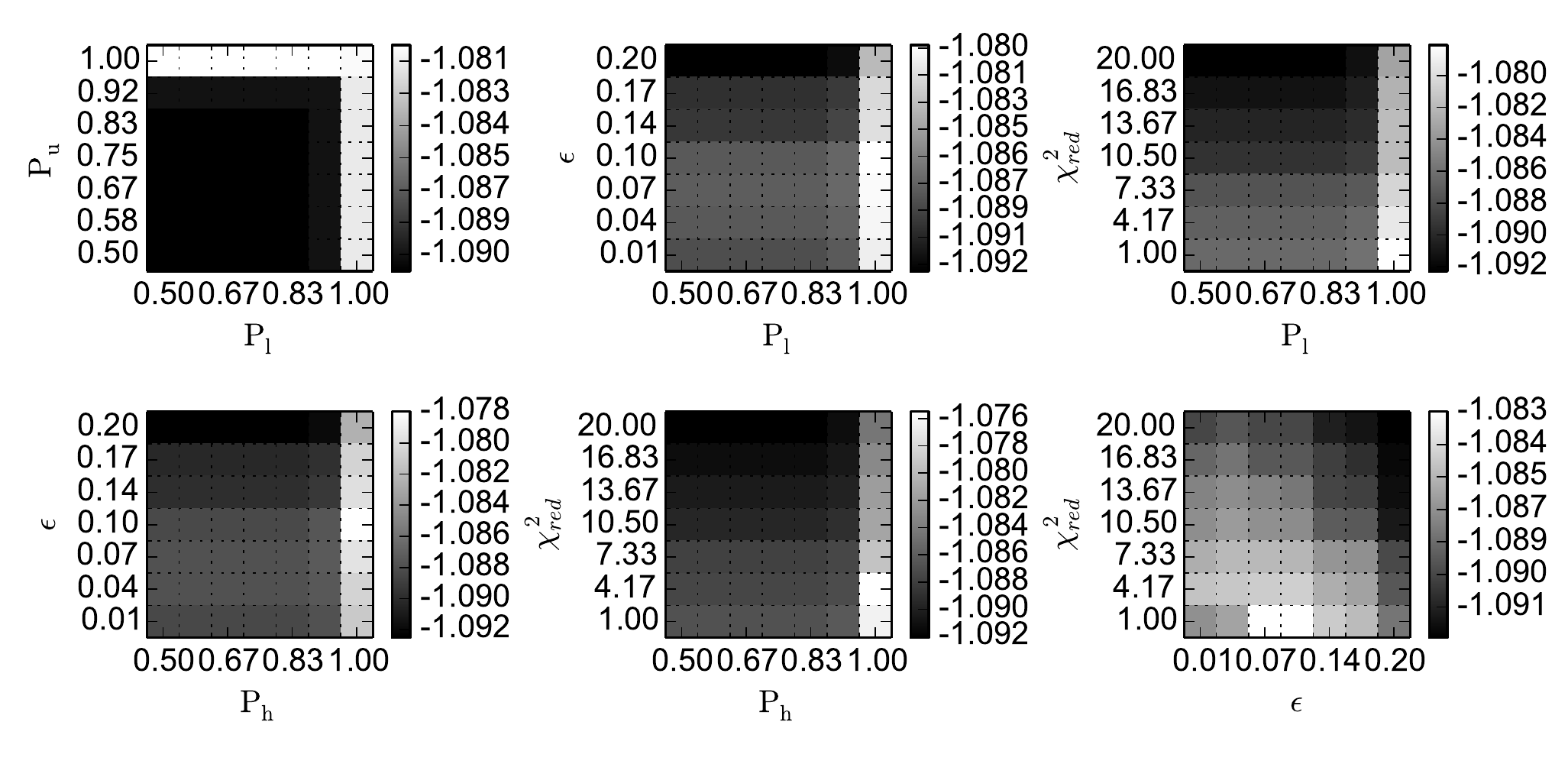}
	\caption{\textsf{The median SI}}
	\label{subfig:SI}
	\end{subfigure}
\caption{\textsf{An exploration of the effects of parameter space for the \texttt{reject} classification and the median of the SI distribution.}}
\label{fig:params-rej-SI}
\end{figure*}

As described in \S\ref{sec:PUMA}, there are 5 parameters that the user declares when running PUMA: the resolution, $\theta_r$; an upper and lower positional probability, $P_u$, $P_l$; two residual fitting metrics, $\chi^2_{red,u}$, $\epsilon_u$. In practice, $\theta_r$ is set by the FWHM of the instrument response of the survey, leaving 4 parameters to be selected by the user. In Figures~\ref{fig:params-iso-dom}-\ref{fig:params-rej-SI}, a range of parameters are run using the mock catalogues created in \S\ref{sec:testPUMA}. The numbers of each PUMA classification, as well as the median SI value, are then plotted as 2D histograms. For each histogram, the mean is taken over the range of parameters not being plotted. The results of Figures~\ref{fig:params-iso-dom}-\ref{fig:params-rej-SI} show that all classifications are robust to the choice of both $P_u$ and $P_l$ with the exception of when $P_u = 1$ or $P_l = 1$., which causes a sharp increase in the number of sources rejected. As intended, as both $\chi^2_{red,u}$ and $\epsilon_u$ increase, the number of accepted sources increases. The choice of these parameters then comes down to the science case of the user, and how important a fit to a power-law model is. The median of the SI distribution is shown to be extremely robust to the choice of parameters.

\end{document}